\newif\ifconfver
\confvertrue       
\ifconfver
\documentclass[10pt,twocolumn,twoside]{IEEEtran}
\else
\documentclass[11pt,draftcls,onecolumn]{IEEEtran}
\fi

\usepackage{makecell}
\usepackage{bbding}
\usepackage{color,graphicx}
\usepackage{float}
\usepackage{multirow,amsmath,epsfig,amsfonts,amssymb,psfig,graphics,psfrag,calc,url,bm,cite}
\usepackage{stfloats,hyperref}
\usepackage{algorithm,algorithmic}
\usepackage{diagbox} 
\usepackage{hhline}
\usepackage{subfigure}
\usepackage{physics}
\usepackage{amsthm}
\usepackage{booktabs}
\usepackage{tikz}
\usepackage{multirow}
\usetikzlibrary{quantikz}
\usepackage{arydshln}
\usepackage{mathtools}

\makeatletter
\def\multilimits@{\bgroup
	\Let@
	\restore@math@cr
	\default@tag
	\baselineskip\fontdimen10 \scriptfont\tw@
	\advance\baselineskip\fontdimen12 \scriptfont\tw@
	\lineskip\thr@@\fontdimen8 \scriptfont\thr@@
	\lineskiplimit\lineskip
	\vbox\bgroup\ialign\bgroup\hfil$\m@th\scriptstyle{##}$\hfil\crcr}
\def\Sb{_\multilimits@}
\def\endSb{\crcr\egroup\egroup\egroup}
\makeatother

	{\end{list}}

\newlength{\twidth}
\ifconfver
\else
\fi
\definecolor{orange}{RGB}{255,107,0}

\newtheorem{Theorem}{Theorem}

{\begin{list}{}{
			\settowidth{\labelwidth}{\mbox{\textnormal{#1}}}%
			\setlength{\leftmargin}{\labelwidth+\labelsep}}}%
	{\end{list}}

\newcommand\bA{\ensuremath{{\bm A}}}
\newcommand\bB{\ensuremath{{\bm B}}}
\newcommand\bC{\ensuremath{{\bm C}}}
\newcommand\bD{\ensuremath{{\bm D}}}
\newcommand\bE{\ensuremath{{\bm E}}}
\newcommand\bF{\ensuremath{{\bm F}}}

\newcommand\bI{\ensuremath{{\bm I}}}

\newcommand\bQ{\ensuremath{{\bm Q}}}

\newcommand\bS{\ensuremath{{\bm S}}}

\newcommand\bU{\ensuremath{{\bm U}}}

\newcommand\bX{\ensuremath{{\bm X}}}
\newcommand\bY{\ensuremath{{\bm Y}}}
\newcommand\bZ{\ensuremath{{\bm Z}}}

\newcommand\ba{\ensuremath{{\bm a}}}

\newcommand\bs{\ensuremath{{\bm s}}}

\newcommand\bv{\ensuremath{{\bm v}}}

\newcommand\bx{\ensuremath{{\bm x}}}

\definecolor{orange}{RGB}{255,107,0}

\ifconfver

\author{Chia-Hsiang Lin,~\IEEEmembership{Senior Member,~IEEE}, Jui-Ting Chen,~\IEEEmembership{Student Member,~IEEE}, \\Zi-Chao Leng,~\IEEEmembership{Student Member,~IEEE}, and Jhao-Ting Lin,~\IEEEmembership{Student Member,~IEEE}}

\title{COS2A: Conversion from Sentinel-2 to AVIRIS Hyperspectral Data Using Interpretable Algorithm With Spectral-Spatial Duality

\thanks{This study was supported by the Emerging Young Scholar Program (namely, the 2030 Cross-Generation Young Scholars Program) of National Science and Technology Council (NSTC), Taiwan, under Grant NSTC 113-2628-E-006-003.
We thank the National Center for Theoretical Sciences (NCTS) and the National Center for High-performance Computing (NCHC) for providing the computing resources.
\textit{(Corresponding Author: Chia-Hsiang Lin)}}
\thanks{C.-H. Lin is with the Department of Electrical Engineering, and with the Miin Wu School of Computing, National Cheng Kung University, Tainan 70101, Taiwan (R.O.C.) 
(e-mail: chiahsiang.steven.lin@gmail.com).}
\thanks{J.-T. Chen is with the Institute of Computer and Communication Engineering, Department of Electrical Engineering, National Cheng Kung University, Tainan, Taiwan (R.O.C.)
(e-mail:  q36134077@gs.ncku.edu.tw).}
\thanks{Z.-C. Leng is with the Institute of Computer and Communication Engineering, Department of Electrical Engineering, National Cheng Kung University, Tainan, Taiwan (R.O.C.)
(e-mail:  q38115558@gs.ncku.edu.tw).}
\thanks{J.-T. Lin is with the Institute of Computer and Communication Engineering, Department of Electrical Engineering, National Cheng Kung University, Tainan, Taiwan (R.O.C.)
(e-mail:  q38091534@gs.ncku.edu.tw).}
}
\else

\fi

\begin{document}
	
	\bibliographystyle{IEEEtran}
	\maketitle
	\ifconfver \else \vspace{-0.5cm}\fi

\begin{abstract}
The Sentinel-2 satellite, launched by the European Space Agency (ESA), offers extensive spatial coverage and has become indispensable in a wide range of remote sensing applications. 
However, it just has 12 spectral bands, making substances/objects identification less effective, not mentioning the varying spatial resolutions (10/20/60 m) across the 12 bands.
%
% insufficient for fine-grained material or object identification. 
%
If such a multi-resolution 12-band image can be computationally converted into a hyperspectral image with uniformly high resolution (i.e., 10 m), it significantly facilitates remote identification tasks.
Though there are some spectral super-resolution methods, they did not address the multi-resolution issue on one hand, and, more seriously, they mostly focused on the CAVE-level hyperspectral image reconstruction (involving only 31 visible bands) on the other hand, greatly limiting their applicability in real-world remote sensing scenarios.
We ambitiously aim to convert Sentinel-2 data directly into NASA's AVIRIS-level hyperspectral image (encompassing up to 172 visible and near-infrared (NIR) bands, after ignoring those absorption/corruption ones).
For the first time, this paper solves this specific super-resolution problem (highly ill-posed), allowing all historical Sentinel-2 data to have their corresponding high-standard AVIRIS counterparts.
We achieve so by customizing a novel algorithm that introduces deep unfolding regularization and $\bQ$-quadratic-norm regularization into the so-called convex/deep (CODE) small-data learning criterion.
Based on the derived spectral-spatial duality, the proposed interpretable COS2A algorithm demonstrates superior spectral super-resolution results across diverse land cover types, as validated through extensive experiments.
\end{abstract}

	
	\ifconfver \else \vspace{-0.0cm}\fi
	
	\ifconfver \else \vspace{-0.5cm}\fi
	
	\ifconfver \else  \fi

\section{Introduction}\label{sec: introduction}

The Sentinel-2 satellite is part of the European Space Agency's (ESA) Copernicus Programme, designed for Earth observation, especially for monitoring land and coastal areas.
Sentinel-2 satellite offers wide spatial coverage, and has become indispensable in various remote sensing applications, including agriculture/crop monitoring, forestry management, water quality observation, land use and land cover change detection, as well as natural disaster monitoring \cite{phiri2020sentinel,segarra2020remote}.
Due to inherent hardware constraints of imaging sensors, coupled with practical considerations such as atmospheric absorption and the requirement for an adequate signal-to-noise ratio (SNR), the spatial resolution of Sentinel-2 imaging systems typically differs across spectral bands.
Specifically, Sentinel-2 data includes four 10-m bands, six 20-m bands, and two 60-m bands.
Note that Sentinel-2 image just has 12 spectral bands, making remote substances/objects identification less effective, not mentioning the varying spatial resolutions across the 12 bands.
Thus, in order to employ the strong identifiability \cite{HISUN,EMI} of hyperspectral imagery to facilitate the subsequent remote sensing missions, such as change detections \cite{hasanlou2018hyperspectral,wu2013subspace} and anomaly detections \cite{HADJT,HADSS}, we are thinking about the possibility of computationally converting such a multi-resolution Sentinel-2 data into hyperspectral data with uniformly high resolution (i.e., 10 m).

Though there are some spectral super-resolution methods \cite{galliani2017learned,he2023spectral}, they did not address the multi-resolution issue on one hand, and, more seriously, they mostly focused on the Harvard/CAVE-level hyperspectral image reconstruction (involving only 31 visible bands) on the other hand \cite{LTRN,CAVEdata}.
This greatly limits their practical applicability in real-world remote sensing scenarios, which often require near-infrared (NIR) bands to have higher atmospheric penetration, to have better material identifiability, as well as to have more effective vegetation observation (e.g., mangrove dynamics) \cite{WOL2017significant,CODEMM}.
Therefore, this work ambitiously aims to convert Sentinel-2 data directly into NASA's AVIRIS-level hyperspectral image \cite{AVIRISdata}, which, after removing those absorption/corruption bands, encompasses up to 172 narrowly-spaced spectral bands contiguously ranging from visible to NIR wavelengths (0.4 to 2.5 $\mu$m).
For the first time, this paper solves this specific super-resolution problem, termed as conversion from Sentinel-2 to AVIRIS (COS2A) problem.
COS2A technique will allow those historical Sentinel-2 data to have their corresponding high-standard AVIRIS counterpart data over the same scene.

Although COS2A is a highly ill-posed inverse problem, it is technically feasible.
In fact, spectral super-resolution has been demonstrated to be feasible not only in imaging software arena but also in the camera hardware implementation \cite{NCCODE}.
Specifically, to meet the strict hardware volume specification in the miniaturized satellites \cite{ferre2022feasibility,AAHCSD}, the flat and nanoscale metamaterials \cite{wang2024advances} have been employed recently to design compact hyperspectral imaging meta-systems, for which a multispectral meta-image is first generated and then spectrally super-resolved to computationally obtain the target hyperspectral imaging results \cite{NCCODE}.
This shows the feasibility of spectral super-resolution even in such a challenging real-world scenario.
Very recently, the quantum deep network (QUEEN) \cite{HyperQUEEN,QUEENG} has also been applied to achieve spectral super-resolution for the highly challenging multispectral unmixing problem, which is an ill-posed and underdetermined source separation task \cite{liutkus2011gaussian}.
Specifically, the QUEEN-based algorithm serves as a virtual quantum light-splitting prism to further split a multispectral band into more hyperspectral bands, thereby achieving the unmixing task through an overdetermined transform with quantum deep image prior (QDIP) \cite{PRIME}.
Nevertheless, existing spectral super-resolution methods have limited spectral upsampling factors less than or just around ten \cite{PRIME,NCCODE,LTRN,MST++}, making them not applicable to the COS2A problem.

Some related spectral super-resolution methods include the multistage spatial-spectral fusion network (MSFN) \cite{MSFN} and reparameterizing coordinate-preserving proximity spectral interaction network (RepCPSI) \cite{RepCPSI}.
MSFN is a U-Net-like convolutional neural network (CNN) designed to learn the spatial correlation and spectral self-similarity, while RepCPSI is a lightweight network utilizing multi-branch convolutional structures during the training phase to facilitate the spatial-spectral feature extraction. 
In RepCPSI, a coordinate-preserving neighboring spectrum-aware attention mechanism is further designed to maintain spatial positional information while capturing local spectral dependencies in a lightweight manner.
Considering the highly ill-posed nature of the target COS2A problem, it is natural to employ the low-rank property of hyperspectral data to mitigate the ill-posedness.
For example, \cite{LTRN} integrates the canonical polyadic (CP) decomposition within a deep network, and enhances global feature modeling through multidimensional attention.
By decomposing high-dimensional features into learnable one-dimensional basis vectors, the method reduces both computational and storage complexities.
For another example, \cite{lin2022fastRGB} integrates the alternating direction method of multipliers (ADMM) and adaptive moment estimation (ADAM) to achieve outstanding spectral super-resolution results via the ADMM-Adam theory, thereby solving the problem under an elegant mathematical scheme.
Nevertheless, all the above methods focus just on the 31-band hyperspectral reconstruction only in the visible spectral range, and thus are not applicable for COS2A.

We should mention the seminal work \cite{MST++}, referred to as multi-stage spectral-wise transformer (MST).
Although it is also a 31-band method, it achieved remarkable performance at the time it was proposed, and won the first place in the renowned NTIRE challenge.
%New Trends in Image Restoration and Enhancement (NTIRE) at CVPR 2022
%
MST is a transformer-based framework that employs spectral-wise attention mechanism by treating each spectral band as an individual token to capture long-range dependencies along the spectral dimension. 
The MST network adopts a multi-stage structure to progressively refine spectral details.
We should also mention some good works that attempted to super-resolve more hyperspectral bands (up to 102 bands), including the spectral-cascade diffusion model (SCDM) \cite{SCDM} and the unfolding spatiospectral super-resolution network (US3RN) \cite{deepUnfolding}.
SCDM adopts a diffusion-based multistage spectral refinement by modeling the spectral probability distribution, while US3RN solves the super-resolution problem via deep unfolding after a tricky reformulation stage.
Although they still do not reach the AVIRIS-level hyperspectral reconstruction, their mechanisms are well developed in our view.

After investigating all the above works, we conclude that AVIRIS-level spectral super-resolution is too challenging to be effectively solved using conventional inverse imaging theories, especially when the AVIRIS data need to be computationally obtained from a multiresolution data, making the COS2A problem even challenging.
Thus, we will employ the so-called convex/deep (CODE) small-data learning theory \cite{CODE}, composed of a convex phase and a deep phase, which was customized for solving challenging hyperspectral inverse imaging problems (e.g., restoration of NASA's highly damaged data) especially under the circumstances of data scarcity.
Note that in COS2A, the training data collection is indeed a daunting and laborious task.
We not only need to identify spatiotemporal overlapping trajectories between AVIRIS and Sentinel-2 sensors, thereby obtaining the training data pairs containing AVIRIS and Sentinel-2 images acquired over the same spatial scene at the same time, but also need to manually conduct several preprocessing stages before the data can be used to meaningfully learn the deep COS2A function (cf. Section \ref{sec: experiment}).

On the other hand, as we do not think the complicated mapping from Sentinel-2 to AVIRIS could be well learned merely using a deep network, we consider the deep solution just as an auxiliary regularization role.
This greatly mitigates the burden of the deep phase in CODE learning, and can be implemented based on the $\bQ$-quadratic-norm regularization scheme in the CODE theory, which will be mathematically introduced in Section \ref{sec:CODE}.
Thus, we will just have a rough deep solution, which, according to the CODE theory, will be fed into a convex optimization algorithm to complete the spectral super-resolution task.
Again, the COS2A super-resolution is too challenging to be directly solved using conventional approaches, so we do not solve the convex optimization phase directly.
Instead, we derive a spectral-spatial duality theorem to transform the highly ill-posed COS2A spectral super-resolution problem into another well-known and more well-studied spatial super-resolution problem (cf. Theorem \ref{thm:SpeSpaDual}), thereby elegantly solving COS2A via simple matrix factorizations.
This is the so-called inverse problem transform \cite{IPT,dehazing}, allowing a hard inverse problem to be reformulated into and solved as a simpler and more well developed inverse problem.
Both our deep and convex phases are interpretable, and the induced COS2A algorithm (i.e., Algorithm \ref{alg:DuQuCODE}) demonstrates superior spectral super-resolution results across diverse land cover types.

The remaining parts of this paper will be organized as follows.
In Section \ref{sec:thm}, the COS2A problem is formulated and solved using CODE theory and spectral-spatial duality.
Data preparation and extensive experimental results are presented in Section \ref{sec: experiment}.
Conclusions are drawn in Section \ref{sec:conclusion}, and mathematical details are given in Appendix \ref{sec:proof-thmSSD}.

\section{The Proposed COS2A Algorithm}\label{sec:thm}

\begin{figure}[t]
\centering
\includegraphics[width=1\linewidth]{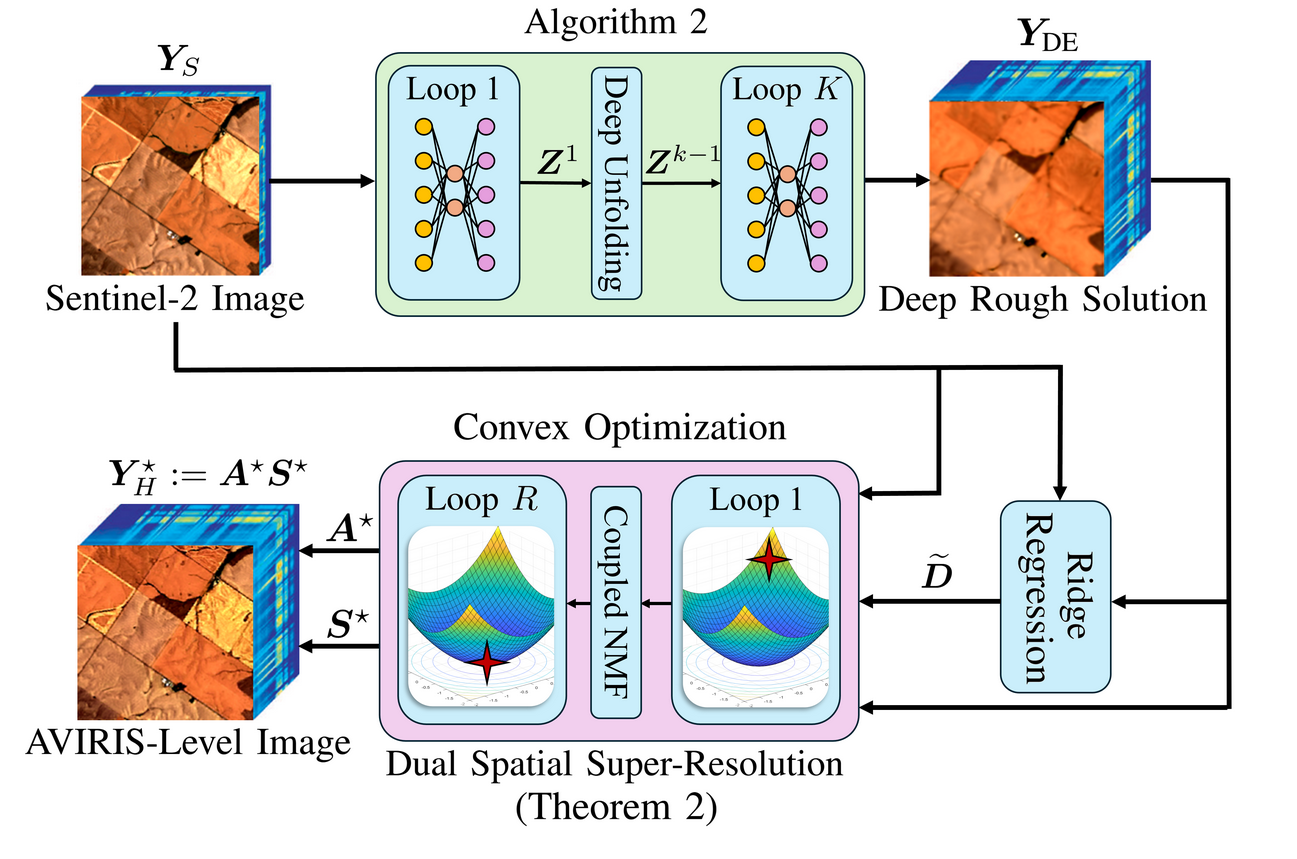}
\vspace{-0.65cm}
\caption{The proposed COS2A spectral super-resolution algorithm (i.e., Algorithm \ref{alg:DuQuCODE}) upgrades the Sentinel-2 multispectral image to AVIRIS-level hyperspectral image. 
It first computes the rough solution $\bY_\textrm{DE}$ using simple deep learning (i.e., Algorithm \ref{alg:ADMM}) and the scene-adaptive spectral response $\widetilde{\bD}$ using ridge regression \eqref{prob:RIDGE}, as a preparation of the subsequent dual spatial super-resolution (cf. Theorem \ref{thm:SpeSpaDual}).
The dual spatial super-resolution is a coupled NMF problem, and is solved using fast convex algorithm \cite{COCNMF}, which returns the reconstructed endmember matrix $\bA^\star$ and abundance matrix $\bS^\star$.
Overall, COS2A judiciously employs the convex/deep learning theory \cite{CODE} to computationally obtain the AVIRIS-level image $\bY_H^\star:=\bA^\star\bS^\star$ with high-fidelity spatial-spectral textures.
}\label{fig:overview}
%\vspace{-0.3cm}
\end{figure}

\subsection{Problem Description and Challenges}\label{sec:probdesc}

Given a multi-resolution Sentinel-2 image $\bY_S\in\mathbb{R}^{12\times L}$, where $L$ is the number of high-resolution pixels in the 10-m band, we aim to computationally reconstruct the corresponding AVIRIS-level hyperspectral image $\bY_H\in\mathbb{R}^{172\times L}$ over the same spatial scene ($M:=172$).
The 12-band Sentinel-2 multi-resolution image can be concisely represented as a matrix $\bY_S$ \cite{SSSS} because those pixels in the 20-m bands (resp., 60-m bands) are copied 4 times (resp., 36 times), as it is when downloaded from ESA's data portal.
Also, although AVIRIS has 224 bands in its original hardware specification, a majority of the literature uses only those uncorrupted bands after removing those absorption/corruption ones, and the remaining 172-band high-quality hyperspectral data (termed as AVIRIS-like data) are proven to be more than sufficient in achieving challenging substances/objects remote identification tasks \cite{HyperKING}.

This inverse problem is highly ill-posed, and regarded as more challenging than typical spectral super-resolution problems that often focus just on reconstructing the 31-band CAVE-level hyperspectral data \cite{RepCPSI}.
Furthermore, it is very difficult to collect high-quality AVIRIS/Sentinel-2 data pairs acquired over the same spatial scene and at the same time, not mentioning the intrinsic non-integer multiple of their spatial resolutions and different flight trajectories/directions \cite{S2data,AVIRISdata}.
This induces a small-data learning problem setting, thereby motivating us to adopt the convex/deep (CODE) learning theory, which was originally developed/customized for restoring NASA's damaged hyperspectral image under the circumstances of data scarcity (and later on applied to numerous satellite monitoring missions) \cite{CODEMM,CODEHCD}.
Simply speaking, CODE avoids complicated mathematics and sophisticated deep learning procedure (e.g., big data collection or architecture design), by using $\bQ$-quadratic-norm regularization to blend the advantages of two powerful inverse imaging techniques---convex optimization (CO) and deep learning (DE).

In Section \ref{sec:CODE}, we outline and recall the overall CODE learning procedure, and formulate the problem of converting Sentinel-2 to AVIRIS data (i.e., COS2A problem) based on the CODE theory \cite{CODE}.
In Section \ref{sec:algdesign}, we design an algorithm to solve the formulated COS2A problem.
The implementation details of our COS2A algorithm are then collectively presented in Section \ref{sec:implement}.

\subsection{CODE-Driven COS2A Problem Formulation}\label{sec:CODE}

Let us begin by recalling the CODE learning theory \cite{CODE}, which was originally proposed for economically solving remote sensing problems without requiring math-heavy optimization or big data collection.
Simply speaking, convex optimization often requires math-heavy regularization (but usually not requiring big data), and deep learning often relies on big data collection and daunting data labeling procedure (but usually not involving advanced mathematics).
CODE learning theory blends the advantages from convex optimization (CO) and deep learning (DE), thereby solving inverse imaging problems without requiring big data or sophisticated regularization.
CODE theory achieved so by bridging CO and DE using the so-called $\bQ$-quadratic-norm regularizer $\|\cdot\|_\bQ$, which not only has very simple mathematic form but also is able to extract useful feature information learned from small data \cite{CODE}.

To be specific, given a target inverse imaging problem, the CODE theory proposes to adopt the following learning criterion to reconstruct the target image $\bY^\star$, i.e.,
\begin{equation}\label{eq:CODEprob}
\bY^\star
:=
\arg\min_{\bY\in\mathcal{Y}}
\textrm{DF}(\bY\mid \bX)+\lambda\textrm{REG}(\bY),
\end{equation}
where $\mathcal{Y}$ often specifies some natural constraints to the target image, $\textrm{DF}(\bY \mid \bX)$ is called the data-fitting term (or data fidelity term) given the observable quantity/data $\bX$, and $\textrm{REG}(\cdot)\triangleq\frac{1}{2}\|\bY-\bY_\textrm{DE}\|_\bQ^2$ is employed as the regularization term with strength controlled by $\lambda\geq 0$.
That said, a pre-computed ``deep'' learning solution $\bY_\textrm{DE}$ is plugged into the ``convex'' $\bQ$-quadratic-norm function for computationally efficient regularization \cite{CODE}, where $\bQ$ is a positive semidefinite (PSD) matrix with the induced norm of a given vector $\bv$ defined as $\|\bv\|_\bQ^2\triangleq \bv^T\bQ\bv$ \cite{CVXbookCLL2016}.

We have three remarks for the CODE regularization scheme, i.e., $\textrm{REG}(\cdot)\triangleq\frac{1}{2}\|\bY-\bY_\textrm{DE}\|_\bQ^2$.
First, $\bY_\textrm{DE}$ could be just a rough deep learning solution \cite{CODE}, meaning that neither a complicated deep network architecture nor a big training dataset is required, greatly mitigating the pressure in developing deep learning algorithms especially when the target inverse problems are highly challenging (e.g., the COS2A problem).
Second, comparing to those commonly used total-variation (TV) regularization that is non-differentiable or self-similarity (SS) regularization that is math-heavy (and usually defined as a non-convex function), the proposed $\bQ$-quadratic-norm regularization is not only differentiable but also has a mathematically very simple form (a concisely defined convex function).
Third, although $\|\cdot\|_\bQ$ is simple, its regularization effectiveness is arguably stronger than those typical TV or SS regularizers \cite[Footnote 1]{CODE}, as $\bY_\textrm{DE}$ often possesses both the edge/similarity structural information embedded in TV/SS (and such information can be extracted by $\|\cdot\|_\bQ$ and fused into the final solution $\bY^\star$).
This has been both theoretically and experimentally verified in \cite{CODE}, thereby extensively inducing numerous successful CODE-based remote sensing applications \cite{CODEIF,NCCODE}.

Accordingly, let us start formulating the COS2A problem based on the CODE theory.
If the spectral response $\bD$ between Sentinel-2 and AVIRIS spectral bands is available, it is natural to define the data fidelity term as $\textrm{DF}(\bY_H \mid \bY_S)=\|\bY_S-\bD\bY_H\|_F^2$.
However, we did not identify an officially released spectral response specification, and hence the matrix $\bD$ should be estimated.
To facilitate the estimation, we focus only on the four highest-resolution bands (i.e., 10-m bands) of the Sentinel-2 image (cf. \eqref{prob:RIDGE}), yielding a reliable spectral response matrix $\widetilde{\bD}\in\mathbb{R}^{4\times 172}$ that will be proven to be sufficient for effectively solving COS2A, and hence the data fidelity term will be designed as $\textrm{DF}(\bY_H \mid \widetilde{\bY}_S)\triangleq \|\widetilde{\bY}_S-\widetilde{\bD}\bY_H\|_F^2$, in which $\widetilde{\bY}_S\in\mathbb{R}^{4\times L}$ represents the concatenation of the four 10-m Sentinel-2 bands.
The estimation criterion of $\widetilde{\bD}$ will be presented in Section \ref{sec:implement}.
Thus, to complete the design of the CODE criterion \eqref{eq:CODEprob} for COS2A, the remaining tasks are twofold---the design of the regularization term and the design of the constraint set $\mathcal{Y}$.

First, the regularization form is defined as those in typical CODE criteria \cite{CODE}, namely $\textrm{REG}(\bY_H)\triangleq\frac{1}{2}\|\bY_H-\bY_\textrm{DE}\|_\bQ^2$, where the deep rough solution $\bY_\textrm{DE}$ will be obtained using simple deep unfolding network, as will be detailed in Section \ref{sec:implement}.
As for the PSD matrix $\bQ$ in the regularizer, it will be customized for $\bY_\textrm{DE}$ to compensate the insufficiency of $\bY_\textrm{DE}$, as will also be detailed in Section \ref{sec:implement}.
Thus, \eqref{eq:CODEprob} leads to the following COS2A formulation, i.e.,
\begin{equation}
\bY_H^\star
:=
\arg\min_{\bY_H\in\mathcal{Y}}
\|\widetilde{\bY}_S-\widetilde{\bD}\bY_H\|_F^2
+
\frac{\lambda}{2}\|\bY_H-\bY_\textrm{DE}\|_\bQ^2,
\end{equation}
where we set $\lambda:=2$ in this work.
The norm $\|\cdot\|_\bQ$ has been previously defined below \eqref{eq:CODEprob} for the vector input, and for the matrix input, the matrix is first vectorized to fit the standard definition \cite{CVXbookCLL2016}.
Second, to specify the constraint set $\mathcal{Y}$, we need to discuss the desired nature of $\bY_H$.
The probably most frequently noticed one would be the low-rank nature of typical hyperspectral image $\bY_H=\bA\bS$, where $\bA\in\mathbb{R}^{172\times N}$ and $\bS\in\mathbb{R}^{N\times L}$ are often considered as the hyperspectral endmember matrix and the abundance matrix, respectively, in which $N$ denotes the matrix rank (often much less than 172 and $L$) \cite{peng2021low,chen2017denoising,SISHY}.
Hierarchically, driven by the low-rank nature of $\bY_H=\bA\bS$, we further employ the minimum-volume nature of $\bA$ \cite{HyperCSI} and the sparsity nature of $\bS$ \cite{giampouras2016simultaneously} to implicitly define $\mathcal{Y}$ at this moment, leading to the COS2A criterion, i.e., 
\begin{equation}\label{prob:CA2SE-AS}
\bY_H^\star
:=
\arg\min_{\bY_H=\bA\bS\in\mathcal{Y}}
\|\widetilde{\bY}_S-\widetilde{\bD}\bA\bS\|_F^2
+
\frac{\lambda}{2}\|\bA\bS-\bY_\textrm{DE}\|_\bQ^2.
\end{equation}
We will discuss more theoretical aspects about the volume and sparsity constraints in Section \ref{sec:algdesign}, thereby making $\mathcal{Y}$ explicit, and accordingly derive the COS2A algorithm for solving \eqref{prob:CA2SE-AS}.

\begin{algorithm}[t]
\caption{COS2A Spectral Super-Resolution Algorithm}
%----------------------------------------------------
\begin{algorithmic}[1]\label{alg:DuQuCODE}
\STATE 
{\bf Given}
Sentinel-2 image $\bY_S$.

\STATE 
Compute the deep rough solution $\bY_\textrm{DE}$ by Algorithm \ref{alg:ADMM}.

\STATE 
Compute the spectral response $\widetilde{\bD}$ by ridge regression using $\bY_\textrm{DE}$ and \eqref{prob:RIDGE}.

\STATE 
Transform the COS2A problem \eqref{prob:CA2SE-exp} (with $\widetilde{\bD}$ and $\bY_\textrm{DE}$) into its dual spatial super-resolution problem via Theorem \ref{thm:SpeSpaDual}.

\STATE 
Conduct the dual spatial super-resolution using fast convex algorithm \cite{COCNMF} to obtain the solution $(\bA^\star,\bS^\star)$ of \eqref{prob:CA2SE-exp}.

\STATE 
{\bf Output} 
AVIRIS-level hyperspectral image $\bY_H^\star:=\bA^\star\bS^\star$.
\end{algorithmic}
%----------------------------------------------------
\end{algorithm}

\subsection{Algorithm for Solving the COS2A Problem}\label{sec:algdesign}

Before solving the COS2A problem \eqref{prob:CA2SE-AS}, we first discuss the theoretical foundation of the volume constraint for $\bA=[\ba_1,\dots,\ba_N]$, where $\ba_i$ is the hyperspectral signature (endmember) of the $i$th underlying substance presented in $\bY_H$.
According to the well known Craig's criterion \cite{craig2002minimum,nascimento2011} in remote sensing area, the high-fidelity endmembers can often be estimated by the vertices of the minimum-volume simplex that encloses the pixel vectors, and under some mild conditions, such an estimation can be proven to yield perfect endmember estimation \cite{EMI,EMIicassp}.
Therefore, we can add an additional simplex volume term $\textrm{vol}(\bA)$ in the objective function to explicitly promote the minimum-volume constraint.
However, simplex volume is known to be non-convex \cite{CVXbookCLL2016,CVXMiniVol}, motivating us to adopt its convex surrogate, which is defined as $\textrm{vol}(\bA)\triangleq\frac{1}{2}\sum_{i=1}^{N-1}\sum_{j=i+1}^N\|\ba_i-\ba_j\|_2^2$.

On the other hand, the sparsity constraint for $\bS$ is oriented from the abundance sparsity nature \cite{iordache2011sparse}, motivating us to add another explicit $\ell_1$-norm term $\|\bS\|_1$ in the objective function \cite{CVXbookCLL2016}.
All in all, criterion \eqref{prob:CA2SE-AS} can be explicitly written as
\begin{equation}\label{prob:CA2SE-exp}
\min_{\bA,\bS\geq \bm 0}
\|\widetilde{\bY}_S-\widetilde{\bD}\bA\bS\|_F^2
+
\frac{\lambda}{2}\|\bA\bS-\bY_\textrm{DE}\|_\bQ^2
+
\alpha \textrm{vol}(\bA)
+
\beta \|\bS\|_1,
\end{equation}
where we further add the non-negative constraints for both endmembers and abundances, and $\alpha,\beta\geq 0$ can be considered as penalty parameters to penalize the violation of the natures discussed in Section \ref{sec:CODE}.
We empirically set $\alpha=\beta=0.002$.
Once we obtain the optimal results $(\bA^\star,\bS^\star)$ from the explicit COS2A criterion \eqref{prob:CA2SE-exp}, we will be able to reconstruct the target image as $\bY_H^\star=\bA^\star\bS^\star$.

Although we can design an alternating optimization algorithm to solve \eqref{prob:CA2SE-exp}, which amounts to two large-scale convex subproblems respectively for solving $\bA$ and $\bS$, we are thinking about solving \eqref{prob:CA2SE-exp} in a more elegant manner, such as by employing some more well-developed super-resolution theory (note that spectral super-resolution is less investigated in literature).
Specifically, as spatial super-resolution methods have been extensively studied in remote sensing and computer vision areas, related theories and methods are considered highly effective \cite{park2003SR}.
Therefore, we are considering to transform the COS2A spectral super-resolution problem into its dual version called spatial super-resolution \cite{park2003SR}, thereby allowing us to elegantly solve the COS2A problem \eqref{prob:CA2SE-exp} by calling effective spatial super-resolution algorithms.
This idea is inspired by the so-called inverse problem transform \cite{IPT,dehazing}---simply speaking, if a tough problem (e.g., convolution) is well transformed into another space (e.g., Fourier-transformed space), it could dramatically become a much simpler problem (e.g., multiplication) that can be elegantly and efficiently solved.

According to the above philosophy, we derive the following theorem based on the non-negative matrix factorization (NMF) theory \cite{NMF}, allowing us to judiciously solve the COS2A problem using well-developed spatial super-resolution theory.
\begin{Theorem}\label{thm:SpeSpaDual}
(The Spectral-Spatial Duality)
The considered COS2A spectral super-resolution problem \eqref{prob:CA2SE-exp} is mathematically equivalent to the well-known coupled-NMF spatial super-resolution problem in optical remote sensing.
\hfill$\square$
\end{Theorem}

\noindent 
To facilitate the readability of the main text, the proof of Theorem \ref{thm:SpeSpaDual} will be given in Appendix \ref{sec:proof-thmSSD}, where one can also find the explicit mathematical definition of the coupled-NMF spatial super-resolution problem commonly seen in the remote sensing area.
Based on Theorem \ref{thm:SpeSpaDual}, we will solve \eqref{prob:CA2SE-exp} by calling a fast convex algorithm \cite{COCNMF} that was customized for the coupled-NMF spatial super-resolution problem.
The overall COS2A algorithm is graphically illustrated in Figure \ref{fig:overview} and summarized in Algorithm \ref{alg:DuQuCODE}, whose implementation details will be given in Section \ref{sec:implement}.

\begin{algorithm}[t]
\caption{Deep Unfolding Algorithm for Solving \eqref{prob:minimize}}
%----------------------------------------------------
\begin{algorithmic}[1]\label{alg:ADMM}
\STATE {\bf Given}
Sentinel-2 image $\bY_S$, and $\rho>0$. 

\STATE 
Initialize $k:=0$ and $\bU^0:=\bm 0$, and spectrally upsample $\bY_S$ to initialize $\bY_H^0$ (cf. Figure \ref{fig:architecture}).

\REPEAT

\STATE 	
Update $\bZ^{k+1}\in\arg\min_{\bZ}~\mathcal{L}(\bY_H^{k},\bZ,\bU^k)$;

\STATE 		
Update $\bY_H^{k+1}\in\arg\min_{\bY_H}~\mathcal{L}(\bY_H,\bZ^{k+1},\bU^k)$;

\STATE 
Update $\bU^{k+1}:=\bU^k-\bY_H^{k+1}+\bZ^{k+1}$;

\STATE 
$k:=k+1$;

\UNTIL the predefined stopping criterion is met.

\STATE {\bf Output} $\bY_\textrm{DE}:=\bZ^k$.
\end{algorithmic}
%----------------------------------------------------
\end{algorithm}

\subsection{Algorithm Derivation and Implementation Details}\label{sec:implement}

Remaining tasks to complete the design of the COS2A algorithm are to explicitly construct the $\bQ$-quadratic-norm regularization, to build a deep network to compute $\bY_\textrm{DE}$, as well as to estimate the spectral response function via $\bY_\textrm{DE}$.
These implementation tasks will be done in this section.

In Algorithm \ref{alg:DuQuCODE}, we need to compute a deep solution $\bY_\textrm{DE}$.
According to the CODE theory \cite{CODE}, the deep solution $\bY_\textrm{DE}$ just needs to be a simple rough solution (thereby avoiding big data or complicated deep network), so we will design a simple deep unfolding network to compute $\bY_\textrm{DE}$ based on the following simple criterion, i.e.,
\begin{equation}\label{prob:DIP}
\bY_\textrm{DE}
:=
\arg\min_{\bY_H}\|\bY_S-\bD\bY_H\|_F^2+\textrm{DIP}(\bY_H),
\end{equation}
where those sophisticated regularization schemes are all merged into the implicit deep regularizer $\textrm{DIP}(\cdot)$, which is theoretically valid based on the seminal deep image prior (DIP) theory \cite{DIP,PRIME}.
The implicit form greatly simplifies the deep network design after unfolding, well aligning with the philosophy of the CODE theory \cite{CODE}.
By the simplified DIP criterion \eqref{prob:DIP}, the associated COS2A deep network will be implemented to convert the Sentinel-2 image $\bY_S$ to the rough AVIRIS-level deep image $\bY_\textrm{DE}$ (cf. Algorithm \ref{alg:ADMM}).

The implicitly regularized simple COS2A criterion \eqref{prob:DIP} can be deeply unfolded using Alternating Direction Method of Multipliers (ADMM) optimizer \cite{CVXbookCLL2016}, thereby implementing the COS2A task using an explicitly constructed deep network.
Specifically, by introducing an auxiliary variable $\bZ\in\mathbb{R}^{M\times L}$, we first reformulate criterion \eqref{prob:DIP} into the standard ADMM form, i.e.,
\begin{equation}\label{prob:minimize}
\min_{\bY_H=\bZ}\|\bY_S-\bD\bY_H\|_F^2+\textrm{DIP}(\bZ).
\end{equation}
The augmented Lagrangian of \eqref{prob:minimize} \cite{CVXbookCLL2016} in the ADMM theory is defined as
\begin{equation}\label{prob:lagrangian}
\mathcal{L}_\rho (\bY_H,\bZ,\bU)
=
\|\bY_S-\bD\bY_H\|_F^2
+\textrm{DIP}(\bZ)
+\frac{\rho}{2} \|\bY_H-\bZ-\bU\|_F^2,
\end{equation}
where $\bU\in\mathbb{R}^{M\times L}$ is the scaled dual variable, and $\rho>0$ is referred to as the penalty parameter in ADMM; $\rho$ is set as a trainable parameter in the deep unfolding network (cf. Figure \ref{fig:architecture}).
Then, the ADMM algorithm \cite{CVXbookCLL2016} for solving \eqref{prob:minimize} via deep unfolding is summarized in Algorithm \ref{alg:ADMM}.

%{\red (output Z or YH?)}

\begin{figure}[t]
\centering
\includegraphics[width=0.9\linewidth]{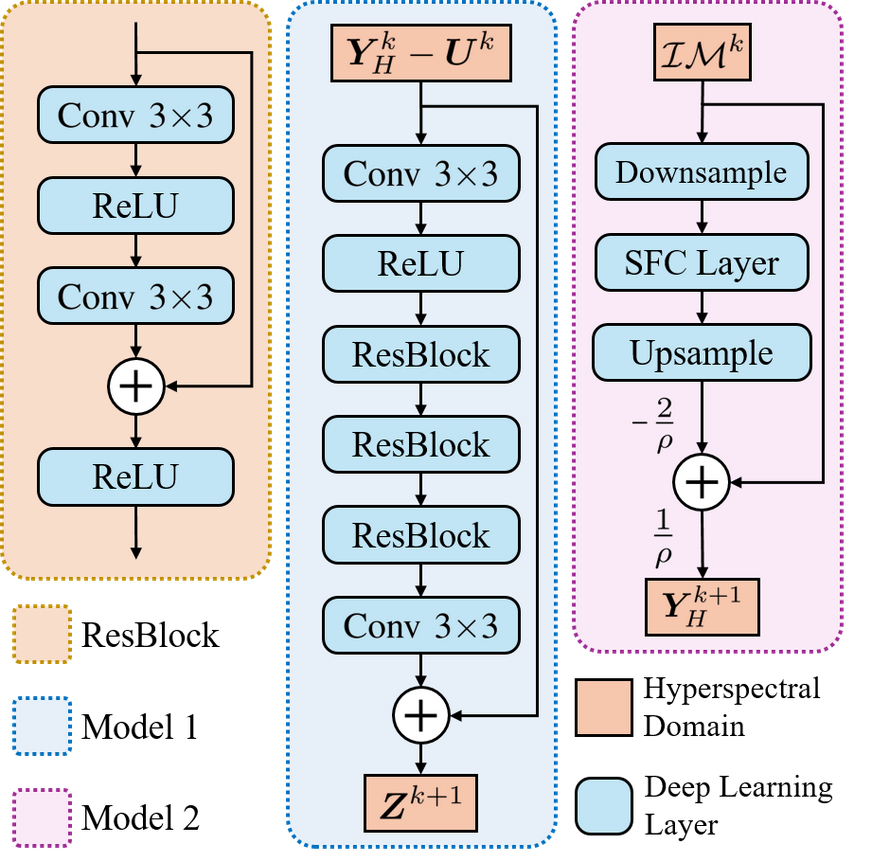}
%\vspace{-0.35cm}
\caption{Model 1 implements the proximal denoiser of \eqref{prob:Z-Update} using the DIP-driven residual-in-residual deep convolutional (Conv) architecture.
Model 2 implements the operator of \eqref{prob:Y_H-Update} based on the symmetricity of $\bm\Phi$ and the physical meanings of $\bD$ (downsampling) and $\bD^T$ (upsampling).
The two models serve as the basic units to deeply unfold the Algorithm \ref{alg:ADMM}.}\label{fig:model12}
%\vspace{-0.3cm}
\end{figure}

To complete Algorithm \ref{alg:ADMM}, we still need to implement the two optimization subproblems for $\bZ^{k+1}$ and $\bY_H^{k+1}$, from which we can deploy the COS2A-driven deep unfolding network.
By the definition of \eqref{prob:lagrangian}, the update of $\bZ$ actually corresponds to the following proximal operator, i.e., 
\begin{align}\label{prob:Z-Update}
\bZ^{k+1}
:=
\textrm{prox}_{\frac{1}{\rho}\textrm{DIP}}(\bY_H^k-\bU^k),
\end{align}
where $\textrm{prox}_{f}(\bv)\triangleq\arg\min_{\bx} f(\bx)+\frac{1}{2}\|\bx-\bv\|_2^2$ denotes the proximal operator associated with a given function $f$ \cite{parikh2014proximal}.
According to this definition of proximal operator $\textrm{prox}_{f}(\bv)$ \cite{parikh2014proximal}, it is nothing but a regularized denoising problem.
Specifically, $\bv$ is considered as a noisy image, from which we aim to recover a clean image $\bx^\star\equiv\textrm{prox}_{f}(\bv)$ with some desired property specified and promoted by the regularizer $f$.
Therefore, the auxiliary hyperspectral image $\bZ^{k+1}$ can be obtained by denoising the image ``$\bY_H^k-\bU^k$'' using the deep network regularizer/prior (i.e., DIP) \cite{DIP,zhang2021plug}.
Accordingly, the deep unfolding network for \eqref{prob:Z-Update} can be deployed as denoising network, for which the residual-in-residual deep architecture is known to be a simple but effective one \cite{zhang2017beyond}, as detailed in the model 1 of Figure \ref{fig:model12}.
The noisy image ``$\bY_H^k-\bU^k$'' can be formed by the architecture deployed in Figure \ref{fig:architecture}, and then be fed into the model 1 to implement \eqref{prob:Z-Update}.

On the other hand, by the definition of \eqref{prob:lagrangian}, the update of $\bY_H$ actually amounts to an unconstrained convex quadratic problem having the following solution, i.e.,
\begin{align*}
\bY_H^{k+1}
&=
\arg\min_{\bY_H}
\|\bY_S-\bD\bY_H\|_F^2+\frac{\rho}{2}\|\bY_H-\bZ^{k+1}-\bU^k\|_F^2
\notag
\\
&=
(2\bD^T\bD+\rho \bI)^{-1}\left(2\bD ^T \bY_S+\rho(\bZ^{k+1}+\bU^k)\right).
\end{align*}
As $\bD$ is a fat matrix, the matrix inversion part in the deep unfolding network is not easily trainable.
To resolve the dilemma, the Woodbury matrix inversion lemma \cite{CVXbookCLL2016} is applied to provide an equivalent solution, i.e.,
\begin{equation}\label{prob:Y_H-Update}
\bY_H^{k+1}
:=
\frac{1}{\rho}\left(\bI - \frac{2}{\rho}\bD^T
{\bm\Phi}
\bD\right)\left(2\bD ^T \bY_S+\rho(\bZ^{k+1}+\bU^k)\right),
\end{equation}
where ${\bm\Phi}\triangleq (\bI+\frac{2}{\rho}\bD\bD^T)^{-1}$ is a symmetric matrix just involving a small-scale matrix inversion.
The input image ``$\mathcal{IM}^k\triangleq 2\bD ^T \bY_S+\rho(\bZ^{k+1}+\bU^k)$'' of \eqref{prob:Y_H-Update} can be formed by the architecture deployed in Figure \ref{fig:architecture}, and then be fed into the model 2 that implements the operator ``$\frac{1}{\rho}\left(\bI - \frac{2}{\rho}\bD^T{\bm\Phi}\bD\right)$'' in \eqref{prob:Y_H-Update}.
Specifically, the identity matrix in the operator is implemented by a shortcut connection, and ``$\bD^T{\bm\Phi}\bD$'' is implemented by concatenating a downsampling unit ($\bD$), a symmetric fully connected (SFC) layer ($\bm\Phi$) and an upsampling unit ($\bD^T$), as depicted in the model 2 of Figure \ref{fig:model12}.

According to the above developed model 1 (for $\bZ^{k+1}$) and model 2 (for $\bY_H^{k+1}$), the deep unfolding network for implementing the $K$-stage Algorithm \ref{alg:ADMM} is deployed in Figure \ref{fig:architecture}.
According to the philosophy of the CODE theory \cite{CODE}, we only need to have a simple rough solution to be used in the $\bQ$-quadratic-norm regularization, so we set $K:=4$ in this work to avoid sophisticated network architecture.
In Figure \ref{fig:architecture}, the $k$th stage ($1<k<K$) can be understood as the preparation of the images ``$\bY_H^k-\bU^k$'' (cf. \eqref{prob:Z-Update}) and ``$2\bD ^T \bY_S+\rho(\bZ^{k+1}+\bU^k)$'' (cf. \eqref{prob:Y_H-Update}), the processing of the prepared images using model 1 (cf. \eqref{prob:Z-Update}) and model 2 (cf. \eqref{prob:Y_H-Update}), as well as the update of the scaled dual variable $\bU^k$ (cf. Algorithm \ref{alg:ADMM}).
Besides, stage 1 has an additional spectral upsampling block serving as the initialization of $\bY_H^0$ (cf. Algorithm \ref{alg:ADMM}), while the last stage (i.e., Stage $K$) is greatly simplified as it just needs to return the target image $\bZ^K$, as illustrated in Figure \ref{fig:architecture}.
Therefore, we are able to obtain $\bY_\textrm{DE}$ through the interpretable deep network designed above.

\begin{figure}[h]
\centering
\includegraphics[width=1\linewidth]{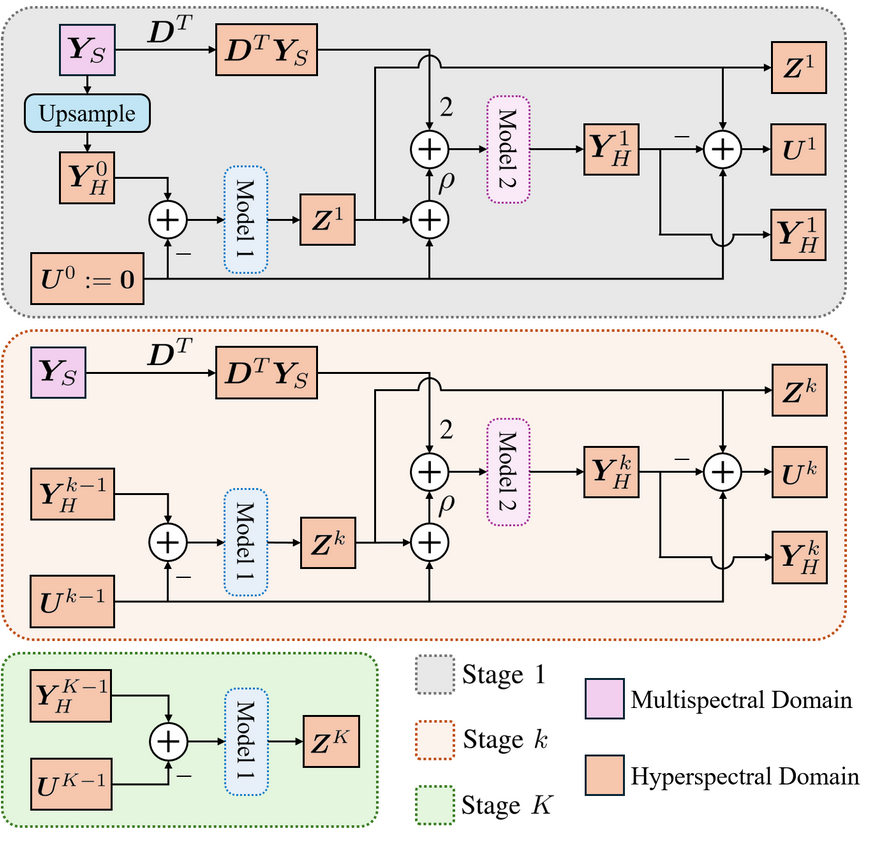}
\vspace{-0.85cm}
\caption{To deeply unfold the $K$-stage Algorithm \ref{alg:ADMM}, our proposed deep architecture implements the operators in \eqref{prob:Z-Update} and \eqref{prob:Y_H-Update} using the unfolded model 1 and model 2 (cf. Figure \ref{fig:model12}), respectively.
This induces a very lightweight architecture (about 0.75M parameters) implemented based on the group convolution.
Simply speaking, the $k$th stage ($1<k<K$) prepares two images [i.e., ``$\bY_H^k-\bU^k$'' (cf. \eqref{prob:Z-Update}) and ``$2\bD ^T \bY_S+\rho(\bZ^{k+1}+\bU^k)$'' (cf. \eqref{prob:Y_H-Update})] to be processed using model 1 (cf. \eqref{prob:Z-Update}) and model 2 (cf. \eqref{prob:Y_H-Update}), respectively, where $\rho$ is a trainable parameter shared in each stage.
The dual variable $\bU^k$ is also updated according to Algorithm \ref{alg:ADMM}.
Stage 1 further implements the initialization of Algorithm \ref{alg:ADMM}, while the last stage (i.e., Stage $K$) is simplified as only the target image $\bZ^K$ is needed as the output of Algorithm \ref{alg:ADMM}.}\label{fig:architecture}
%\vspace{-0.3cm}
\end{figure}

%(in deep unfolding $\lambda=\eta$...)
%($\eta$ is removed, so $\eta=1$...)

As the regularizer $\|\cdot\|_\bQ$ is interpreted in the CODE theory as a simultaneous feature extraction and fusion operator \cite[Equation (3)]{CODE}, the design of $\bQ$ aims to extract useful information from $\bY_\textrm{DE}$ and fuses the information into the target image $\bY_H$.
The desired information from $\bY_\textrm{DE}$ is obviously its spectral shape.
To utilize the spectral information of $\bY_\textrm{DE}$, there are two potential approaches, corresponding to the pixel-level and region-level spectral information usages.
Since the deep unfolding criterion \eqref{prob:DIP} accepts the multi-resolution input $\bY_S$ (including the low/medium-resolution bands), making $\bY_\textrm{DE}$ appears to be somewhat blurred.
That said, pixel-level approach would not be effective, and thus we apply the region-level approach that forces neighboring pixels/spectra to cooperate with each other.
Specifically, an $r\times r$ region integration (judiciously modeled by some blurring kernel $\bB$) is first performed on $\bY_\textrm{DE}$, before it is used to $\bQ$-regularize the ill-posed COS2A problem, thereby leading to
\begin{equation}\label{def:Q}
\bQ
\triangleq 
\bB\bB^T\otimes\bI_M\in\mathbb{R}^{(ML)\times (ML)},
\end{equation}
where $\otimes$ denotes the Kronecker product.
Also, as the blurring effect comes from the low/medium-resolution input bands of $\bY_S$ (the 60-m bands has a blurring factor 6, while the 20-m bands has a blurring factor 2), it is natural to model the kernel to have a blurring factor $r:=2$ as Sentinel-2 is dominated by the 20-m bands (there are six 20-m bands, while there are just two 60-m bands).
Moreover, as those pixels in the 20-m bands (resp., 60-m bands) are simply copied 4 times (resp., 36 times) as mentioned in Section \ref{sec:probdesc}, this amounts to the so-called uniform blurring, which has the mathematical form of $\bB\triangleq \bI_{L/r^2}\otimes (\bm 1_{r^2}/r^2)\in \mathbb{R}^{L\times (L/r^2)}$ \cite{COCNMF}, where $\bI_n$ and $\bm 1_n$ denote the $n\times n$ identity matrix and the $n$-dimensional all-one vector, respectively.

% (implementation trick) instead of using the information of $\bY_\textrm{DE}$ to estimate $\bY_H$ directly, we inject the ``blurred version of $\bY_\textrm{DE}$'' into ``blurred version of $\bY_H$'' when estimating $\bY_H$ {\red (by doing so, some estimation errors (or the blurring errors) existed in $\bY_\textrm{DE}$ would be somewhat filtered out...)}..............therefore, it can be ``modeled'' as minimization of $\red \|\bY_\textrm{DE}\bB - \bY_H \bB\|_F$ for some blurring kernel $\bB$... 

The remaining task is to present the implementation details about the spectral response matrix $\bD$.
We had ever tried to learn a universal $\bD$ from sufficiently large amounts of real AVIRIS and Sentinel-2 data, but it turned out to be less effective probably due to the spatially invariant attributes.
A theoretically better approach is hence to adaptively learn the spectral response for each data.
To simplify this challenging task, we recall Theorem \ref{thm:SpeSpaDual}, which suggests that the estimated spectral response will be simply used in the dual spatial super-resolution stage.
That said, we can just focus on learning the spectral response $\widetilde{\bD}$ for those high-spatial-resolution bands (i.e., 10-m bands) in $\widetilde{\bY}_S$ \cite{SSSS}.
To avoid the responses being concentrated in limited hyperspectral bands, we adopt the ridge regression \cite{mcdonald2009ridge}, which promotes uniformly distributed solutions, thereby leading to the estimation criterion, i.e.,
\begin{equation}\label{prob:RIDGE}
\widetilde{\bD}
:=
\arg\min_{\bD\geq\bm 0}
\|\bD\bY_\textrm{DE}-\widetilde{\bY}_S\|_F^2
+\eta\|\bD\|_F^2, 
\end{equation}
where $\eta:=0.0001$ is an empirical setting, and $\bY_\textrm{DE}$ is used as a surrogate of the real AVIRIS data; note that in practice, we do not have real AVIRIS data in the COS2A problem.

We have completed the design of the COS2A algorithm (i.e., Algorithm \ref{alg:DuQuCODE}) based on the CODE theory \cite{CODE} to solve the highly ill-posed COS2A super-resolution problem, allowing historical Sentinel-2 data being computationally converted into their corresponding high-standard AVIRIS counterpart.
For the first time, the COS2A super-resolution is effectively achieved  via the interpretable deep network (i.e., Algorithm \ref{alg:ADMM}) and the spectral-spatial duality (i.e., Theorem \ref{thm:SpeSpaDual}). 
COS2A has been achieved for the first time, and its effectiveness will be experimentally demonstrated next.

%RMK:
%CO-CNMF($\mathbf{Y}_h$,$\mathbf{Y}_m$,$\mathbf{B}$,$\mathbf{D}$)
%=
%CO-CNMF($\bY_\textrm{DE}\bB$,$\widetilde{\bY}_S$,$\bB$,$\widetilde{\bD}$)

\section{Experimental Results and Discussion}\label{sec: experiment}

This section comprehensively evaluates the proposed COS2A algorithm (i.e., Algorithm \ref{alg:DuQuCODE}), demonstrating its effectiveness.
The experimental design and dataset description are provided in Section \ref{sec:experimental setting}, where one can find a data preparation protocol customized for the COS2A problem.
As there is no prior COS2A method in existing literature, Section \ref{sec:Results} begins by reporting the baseline preparation, followed by the discussion about the qualitative and quantitative testing results.
%
%It is worth noting that the real-world data collection for COS2A is a non-trivial and labor-intensive procedure.
%
%The procedure for acquiring AVIRIS and Sentinel-2 data, along with multiple preprocessing steps, is detailed in Section \ref{sec: real data sop}.
%
In Section \ref{sec: real result}, we demonstrate the real-world applicability of the proposed COS2A method, directly working on real Sentinel-2 data, thereby yielding high-fidelity hyperspectral reconstruction results that hold good resemblance to real AVIRIS data over the same scene. 

\subsection{Experimental Setting}\label{sec:experimental setting}

Since prior works never consider the COS2A problem, we can not find existing experimental protocol or open training dataset for performance evaluation.
In view of this, we propose a simulation protocol for fair quantitative evaluation, as well as a real data collection protocol for demonstrating real-world applicability of our method.

We first collect 646 AVIRIS hyperspectral images \cite{AVIRISdata}, each with a 256 $\times$ 256 spatial size, available from the AVIRIS dataset offered in \cite[Table 1]{DCSN}, and the ratio of training/validation/testing is 15:1:1.
After removing water-vapor absorption bands (i.e., bands 1–10, 104–116, 152–170, and 215–224) \cite{HyperQUEEN, HyperKING} from the original 224 AVIRIS bands (covering wavelengths from about 400 nm to 2500 nm), the 172-band AVIRIS data $\bY_H$ are obtained and used in the experiments.
To simulate the spectral and spatial characteristics of Sentinel-2 data, we first design a spectral response matrix $\bD\in\mathbb{R}^{12\times 172}$ that maps the AVIRIS spectra to the 12 Sentinel-2 bands (after excluding the low-resolution band 10, which basically records only the cirrus information \cite{SSSS}).
Specifically, according to the central wavelength and the bandwidth of each Sentinel-2 band \cite{S2data,SSSS}, we can define its wavelength coverage, and accordingly identify the subset of the AVIRIS bands corresponding to that Sentinel-2 band; as we did not find a public spectral response between AVIRIS and Sentinel-2 sensors, AVIRIS bands in the subset are averaged to form the simulated Sentinel-2 band.
Then, among the 12 simulated Sentinel-2 bands, those with 10-m band indices (i.e., 2, 3, 4, and 8) bands are kept unchanged to simulate real high-resolution Sentinel-2 bands, while those with 20-m band indices (i.e., 5, 6, 7, 9, 11, and 12) and those with 60-m band indices (i.e., 1, and 10) are uniformly blurred by factors of 2 and 6, respectively, resulting in a simulated multi-resolution Sentinel-2 image $\bY_S$.
The above procedure provides AVIRIS/Sentinel-2 training data pairs ($\bY_H,\bY_S$) with ground-truth $\bY_H$ for quantitatively evaluating the testing results.

On the other hand, to demonstrate the real-world applicability of the proposed COS2A algorithm, we also try hard to identify spatially and temporally overlapping AVIRIS/Sentinel-2 data pairs, so that the quality of the spectrally super-resolved Sentinel-2 data (obtained via Algorithm \ref{alg:DuQuCODE}) can be evaluated based on the real AVIRIS data.
However, as the flying trajectories of AVIRIS/Sentinel-2 sensors are not aligned, image rotation calibration is also needed as a preprocessing stage, besides addressing the resolution inconsistancy between the two sensors.
Specifically, we first download the AVIRIS Level-2 data (reflectance data) from \cite{AVIRISdata} and Sentinel-2 Level-2A data from \cite{S2data}.
Then, the downloaded AVIRIS images are resampled to a 10-m spatial resolution using the benchmark ENVI software \cite{ENVI} in order to meet the highest resolution of Sentinel-2 data, and, as aforementioned, only 172 high quality AVIRIS bands are used \cite{HyperQUEEN,HyperKING}.
Also, for Sentinel-2 data, the pixels in those 20-m bands (resp., 60-m bands) are already copied 4 times (resp., 36 times) when they are downloaded \cite{S2data}, so they are directly rotated to the direction aligning with the AVIRIS sensor trajectory; this orientation calibration can be done using the trajectory angle information provided in the AVIRIS metadata as well as the ENVI software \cite{ENVI}.
We also conduct the image registration between AVIRIS/Sentinel-2 data using ENVI \cite{ENVI} to ensure accurate spatial alignment, thereby facilitating the learning of deep COS2A function.
ENVI software is also used to remove irregular black borders and cloud-covered regions from the data, and the same cropping areas are applied to both AVIRIS/Sentinel-2 data to maintain their spatial consistency, thereby resulting in a total of 526 real-data pairs, each composed of a 256$\times$256$\times$172 AVIRIS image and a 256$\times$256$\times$12 Sentinel-2 image acquired over the same scene at the same time.
The data are divided for training/validation/testing with the ratio of 8:1:1.
Other related real data information, including the acquisition sites, locations, ground sampling distances (GSD), spectral ranges and spatial sizes, are summarized in Table \ref{tab:data_combined}.

% Real Data ROI---------------------------
\begin{table*}[t]
\centering
\caption{Real AVIRIS and Sentinel-2 data pairs are downloaded from \cite{AVIRISdata} and \cite{S2data}, respectively. 
All sites are located in the America, and all data were acquired in August 2019. 
Sentinel-2 spectral ranges are covered by the AVIRIS sensors, and Sentinel-2 image sizes are presented with respect to the 10-m bands.}
\label{tab:data_combined}
\scalebox{0.95}{
\begin{tabular}{|c|c|c|c|c|c|c|c|}
\hline
\multirow{2}{*}{Site Name} & \multirow{2}{*}{Location} & \multicolumn{3}{c|}{AVIRIS} &  \multicolumn{3}{c|}{Sentinel-2} \\
\cline{3-8}
& & GSD (m) & Spectral Range ($\mu$m) & Image Size & GSD (m) & Spectral Range ($\mu$m) & Image Size \\
\hline
Mill Fire 2 &  CA  & 16 & 0.4–2.5 & 8544$\times$803 & 10 / 20 / 60 & 0.443–2.19 & 13670$\times$1285 \\
Western Sierra Foothills & CA  & 14.8 & 0.4–2.5 & 10486$\times$855 & 10 / 20 / 60 & 0.443–2.19 & 15519$\times$1265 \\
Caltech & CA & 16.1 & 0.4–2.5 & 9319$\times$810 & 10 / 20 / 60 & 0.443–2.19 & 15004$\times$1304 \\
Williams Fire (Rose) & WA & 16.1 & 0.4–2.5 & 9793$\times$782 & 10 / 20 / 60 & 0.443–2.19 & 15767$\times$1259 \\
\hline
\end{tabular}
}
\end{table*}

For the training setting of the proposed COS2A algorithm, the batch size is set to 8, and the deep unfolding network (cf. Figure \ref{fig:architecture}) is trained for 30 epochs. 
%
% The network parameters are optimized using the ADAM optimizer \cite{adam} with an initial learning rate of 1E-4, which is reduced by a factor of 0.5 every 30 epochs using a multistep learning rate scheduler.
%
A total of 20000 overlapped 64$\times$64 patches are cropped for training and validation.
Data augmentation methods, such as random horizontal/vertical flips and rotation, are applied during the training phase.
The COS2A model is trained in an end-to-end manner by minimizing the outlier-robust $\ell_1$ loss function \cite{CODE}, i.e., $\|\bY_H-\bY_\textrm{DE} \|_1$, where $\bY_H$ is the reference image and $\bY_{\text{DE}}$ is the network output during the training stage.
The spectral response matrix $\bD$ is known as discussed above, so it is applied for the simulation study, while it should be estimated via ridge regression for the real data experiment.
Besides, the model-order $N$ is empirically set as 10 \cite{COCNMF}, which yields good performance in general.
%
%For MST, we adopt the training setting of {\red 30} epochs \cite{MST++}.
%
All other training configurations, including the optimizer, learning rate scheduling, and data augmentation strategies, are kept identical to those used for COS2A to ensure a fair comparison.
All the experiments are conducted under MATLAB R2023b, Python 3.12.3/3.12.4, and PyTorch 2.6.0/2.7.0 on a computer equipped with an Intel Core i9-10900K CPU with 3.70-GHz speed, and 46-GB of system RAM, and an NVIDIA GeForce RTX 3090 GPU.

\subsection{Qualitative and Quantitative Analysis}\label{sec:Results} 

As this is the first work that achieves the COS2A function, which computationally upgrades the 12-band multi-resolution Sentinel-2 data to 172-band AVIRIS-level hyperspectral data, we have to design a preliminary baseline for comparison.
As reported in Section \ref{sec: introduction}, almost all the related works focus on reconstructing 31-band hyperspectral images from 3-band RGB images.
Thus, it is natural to introduce the divide-and-conquer (DAC) algorithm \cite{DACalgo} to prepare for the baseline.
Specifically, as COS2A is a 12-to-172 super-resolution function, we divide the function into four parallel blocks, each conducting a 3-to-43 spectral super-resolution task.
However, as most related works are for 3-to-31 super-resolution, they should be augmented with an additional fully-connected layer to convert 31 feature maps into 43 ones.
After each of the four blocks returns 43 features, they are aggregated into a 172-band hyperspectral image, which will be used as the baseline for our COS2A function.
Each block here can be implemented using any CAVE-level method, and here we opt to use the seminal MST algorithm \cite{MST++}, which won the champion in the NTIRE 2022 spectral reconstruction challenge \cite{arad2022ntire}.
Note that although we have ever tried to directly modify the MST algorithm to fit the COS2A problem, it did not come up with an acceptable results before adopting the DAC strategy (even after nearly 3-week training period).
The reason is that ever for such an outstanding work \cite{MST++}, its configuration cannot handle the highly challenging COS2A mission, and that is why we have to introduce the DAC strategy to divide the COS2A problem into some simpler manageable subproblems.
The above DAC configuration is trained over 30 epochs for each branch, enabling each branch to specialize in a specific and manageable portion of the spectrum learning, thereby effectively completing the DAC baseline design.

\begin{figure}[t]
    \centerline{\includegraphics[width=0.5\textwidth]{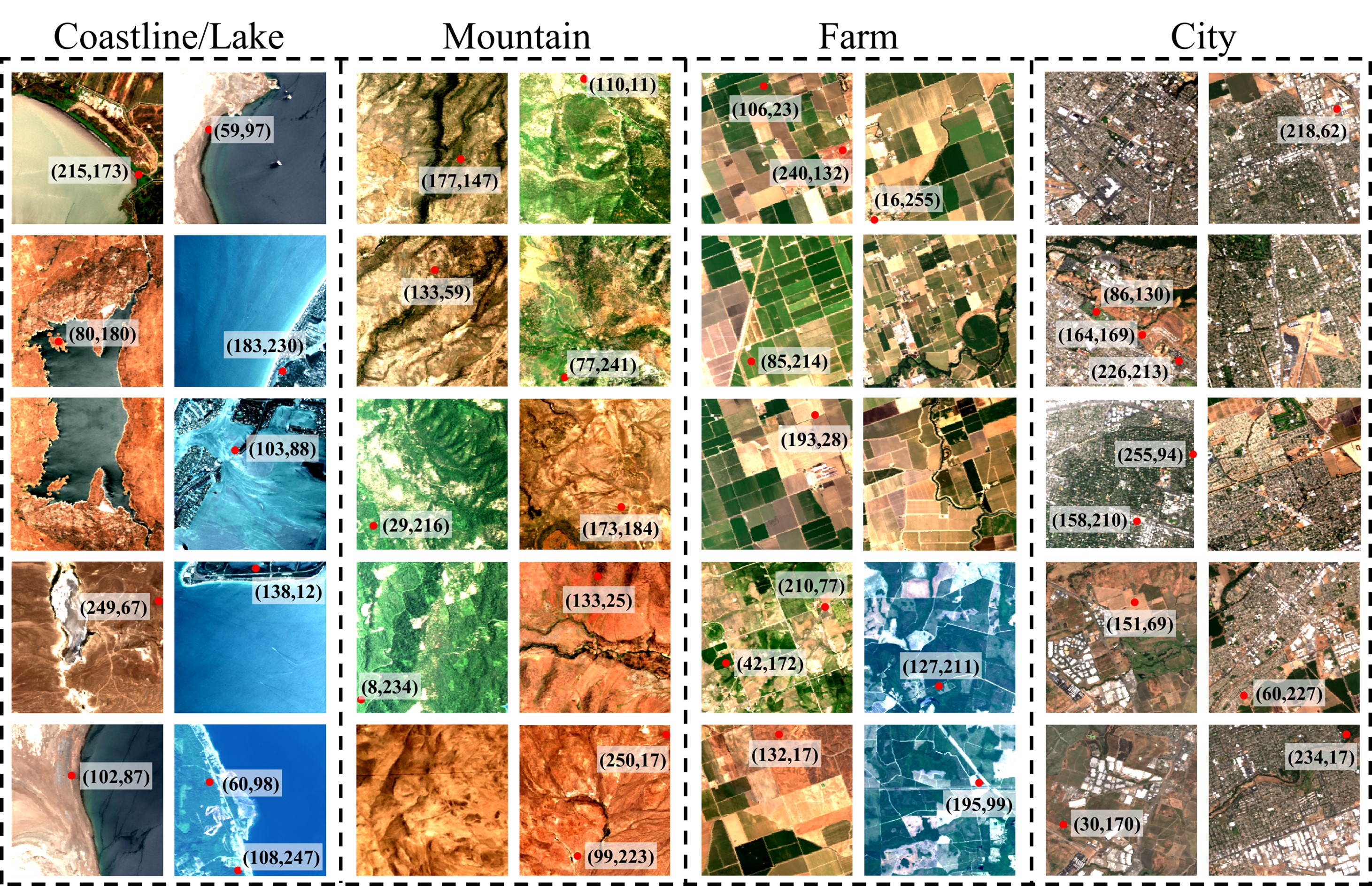}}
    \caption{The real AVIRIS images over the four representative land types, including coastline/lake, mountain, farm, and city.
    For each land type, we also select ten representative pixels (spectrally distinct pixels) as indicated by the red dots for the subsequent evaluation of spectral reconstruction quality (cf. Figure \ref{fig: spectral}).
    }\label{fig: 40}
\end{figure}

\begin{figure}[t]
    \centerline{\includegraphics[width=0.5\textwidth]{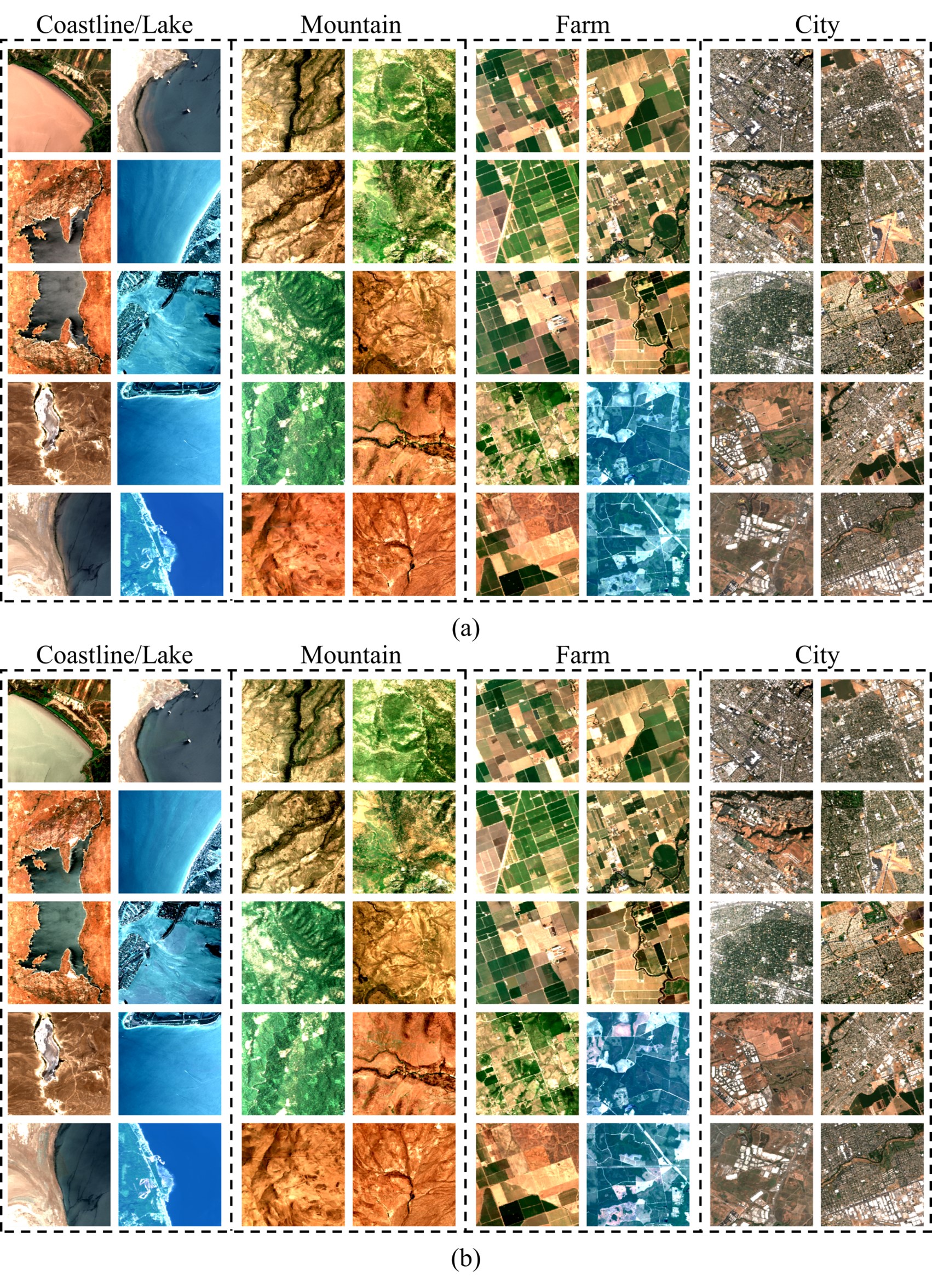}}
    \caption{Spatial reconstruction results of (a) DAC and (b) the proposed COS2A algorithm over coastline/lake, mountain, farm, and city.
    All the images used are displayed in the true-color composition [B23(r)-B12(g)-B5(b)].
    }\label{fig: 40_results}
\end{figure}

For effectively evaluating the similarity between COS2A-generated $\bY_H^\star$ and the real AVIRIS image $\bY_H$, we perform illumination calibration between the AVIRIS image and Sentinel-2 data (cf. Section \ref{sec: real result}).
We demonstrate the generalization ability of the COS2A across diverse scenes by categorizing the 40 testing data into four representative land types, namely coastline/lake, mountain, farm, and city.
Each land type includes 10 data, and images for these four land types are displayed in Figure \ref{fig: 40}.
The visual reconstruction performance is presented in Figures \ref{fig: 40_results}.

Both DAC and the proposed COS2A algorithm exhibit outstanding spatial reconstruction results across most of the 40 data.
However, some noticeable distortions are observed in the results of DAC.
For example, in the coastline/lake type, the waterbody region in the first image of the left column shows a serious color deviation. 
Similarly, for the mountain type, the second image of the right column suffers from color inconsistency in the right half region. 
For the farm type, the bright regions in the fourth and fifth images of the right column are also discolored.
In contrast, the proposed COS2A algorithm not only preserves fine spatial details across all the land types, but also maintains color fidelity more consistently, alluding the spectral fidelity of the COS2A-reconstructed images as will be further validated next.

\begin{figure}[t]
    \centerline{\includegraphics[width=0.5\textwidth]{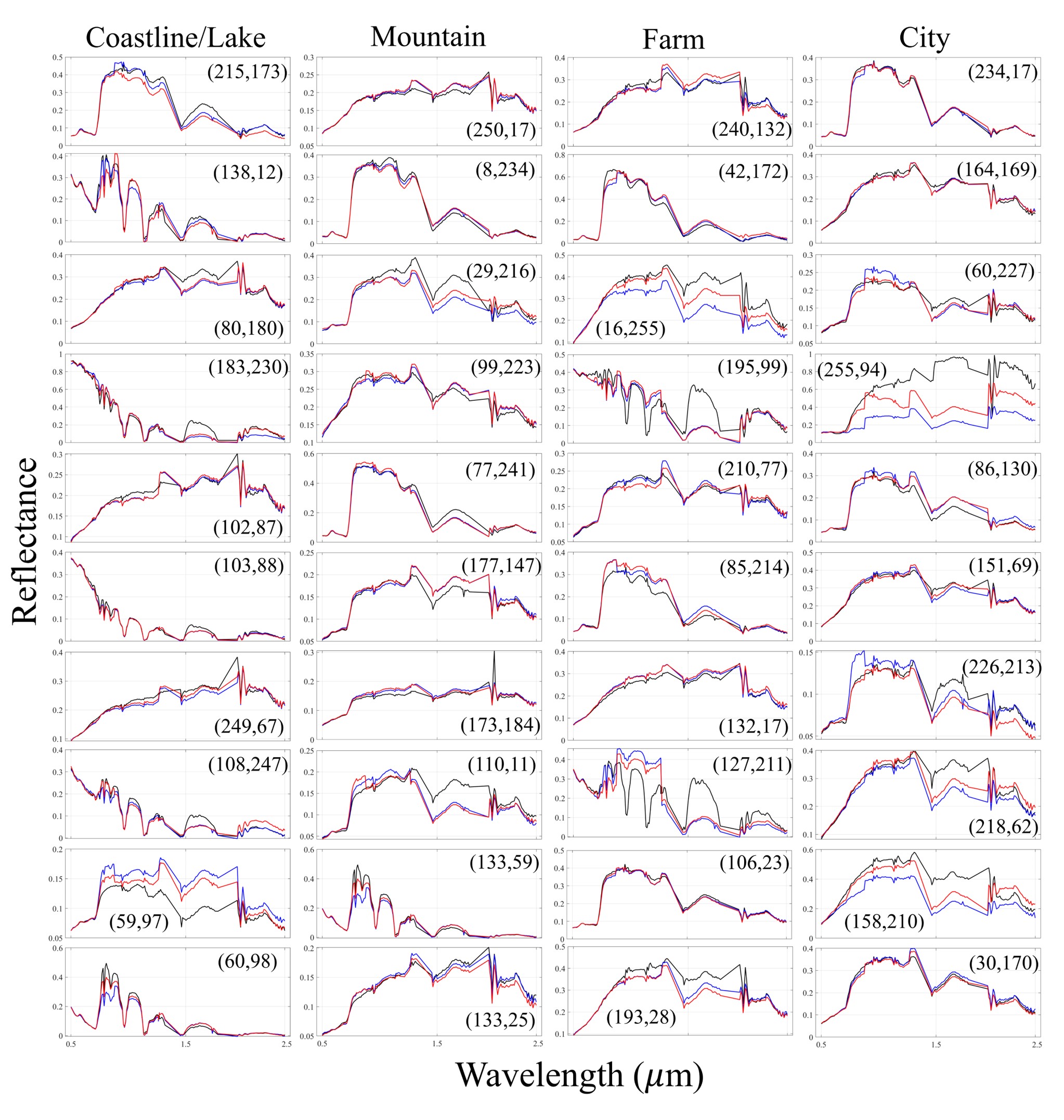}}
    \caption{Spectral reconstruction results of (a) DAC and (b) the proposed COS2A algorithm over coastline/lake, mountain, farm, and city.
    The DAC-reconstructed and COS2A-reconstructed hyperspectral pixels are displayed as blue and red curves, respectively, while the real AVIRIS pixels are displayed as black curves.
    The locations of the selected pixels are marked as red dots in Figure \ref{fig: 40_results}.
    }\label{fig: spectral}
\end{figure}

Since we test 40 images in total, each with 65536 pixels, it is impractical to display all the pixel curves due to the page limitation.
%for multiple pixels from each image.
%
Thus, the ten images in each land type are jointly fed into the successive projection algorithm \cite{SPA} to collaboratively identify ten representative pixels (i.e., spectrally distinct pixels) for each land type, as displayed in Figure \ref{fig: 40}.
The spectral reconstruction results for these selected pixels are collectively presented in Figure \ref{fig: spectral}.
For coastline/lake area, the pixel at (108,247) demonstrates that the COS2A-reconstructed hyperspectral pixel highly overlaps with the real AVIRIS pixel, especially capturing the three continuous appeared peaks in the range of 0.7-0.95 $\mu$m.
Remarkably, even for a more challenging pixel such as (59,97), where both DAC and COS2A show noticeable deviations, the reconstructed pixels from both methods still preserve the overall spectral shape and remain similar to the real AVIRIS pixel.
In the mountain area, the pixels at (99,223) and (133,25) show a high degree of alignment between the COS2A-reconstructed pixel and the AVIRIS one.
For other pixels, such as (29,216) and (117,147), while slight deviations are observed, both DAC and COS2A are able to preserve the overall spectral shape.
In particular, both methods are able to capture the sharp oscillation patterns in the NIR spectral range.
For farm area, the pixel at (127,211) appears to be a more challenging case for spectral reconstruction. 
Nevertheless, COS2A still generates a spectral curve that more closely follows the ground-truth one than DAC, highlighting its robustness under challenging conditions.
The pixel at (16,255) demonstrates that both methods achieve accurate spectral shapes, but COS2A again shows a much closer match to the real AVIRIS pixel, indicating its superior spectral fidelity.
For city area, which is generally considered a more complex area for spectral reconstruction due to its heterogeneous structures.
We observe that the pixels, such as (255,94) and (158,210), exhibit more distinctive spectral characteristics.
Even under these challenging conditions, COS2A consistently reconstructs spectra that are closer to the real AVIRIS pixels than DAC, while still preserving a high spectral similarity.
In addition, COS2A achieves excellent reconstruction results at pixel (30,170), further demonstrating its effectiveness in urban areas.
Overall, COS2A-reconstructed pixels demonstrate highly consistent spectral shapes with real AVIRIS pixels (cf. Figure \ref{fig: spectral}), suggesting that it does effectively capture the fine characteristics of AVIRIS pixel features hence leading to faithful spectral reconstruction results.

To comprehensively and fairly evaluate the performance of our method, we need quantitatively analyze the results.
Specifically, we adopt peak signal-to-noise ratio (PSNR) \cite{CODEIF}, spectral angle mapper (SAM) \cite{SISHY}, root mean squared error (RMSE) \cite{DCSN}, and structural similarity (SSIM) \cite{CODE} for quantitative evaluation.
The results averaged over the testing data are summarized in Table \ref{tab: quan} with the best performance of each scenario marked by a boldfaced number, where one can see that the spatial reconstruction quality achieved by COS2A is much superior than DAC as indicated by the PSNR values.
As for the SAM index, which is the most important measure for the spectral shape preservation, COS2A significantly leads DAC by several degrees.
Although DAC has reasonable spectral reconstruction results (less than ten degrees) \cite{HyperCSI}, the COS2A-generated hyperspectral pixels show much better resemblance to the real AVIRIS hyperspectral pixels, as quantitatively validated in Table \ref{tab: quan}.
This well echoes the qualitative analysis in Figure \ref{fig: spectral}, although COS2A takes longer (but reasonable) time to achieve so.
RMSE is to measure global reconstruction quality.
Given the outstanding spatial and spectral reconstruction quality (cf. PSNR and SAM indices in Table \ref{tab: quan}), it not surprising that COS2A achieves very low global reconstruction error as indicated by RMSE.
Finally, as can be seen from the superior SSIM performance (cf. Table \ref{tab: quan}), the COS2A-generated images do preserve outstanding structural similarity to the real AVIRIS images.
This can also be observed by comparing the structural properties between the images displayed in Figure \ref{fig: 40} and Figure \ref{fig: 40_results}(b).

\begin{table}[t]
\footnotesize
\centering
\caption{Quantitative evaluation of the COS2A performance. 
The boldfaced numbers indicate the best performances (i.e., the largest PSNR/SSIM, or the smallest SAM/RMSE).}
\label{tab: quan}
\setlength{\tabcolsep}{1mm}
\renewcommand{\arraystretch}{1.2}
\scalebox{0.94}{
\begin{tabular}{c|c|ccccc}
\hline
& Methods & PSNR $\!(\uparrow)$ & SAM $\!(\downarrow)$ & RMSE $\!(\downarrow)$ & SSIM $\!(\uparrow)$ & Time (sec.) \\
\hline
\multirow{2}{*}{\makecell{Coastline/\\Lake}}
& DAC & 27.6972 & 8.4138 & 0.0264 & 0.9285 & \bf{0.0665} \\
\cdashline{2-7}
& COS2A & {\bf35.8897} & {\bf2.4539} & {\bf0.0080} & {\bf0.9704} & 15.3353 \\
\hline
\multirow{2}{*}{\makecell{Mountain}}
& DAC & 28.2909 & 4.9266 & 0.0180 & 0.9317 & \bf{0.0659} \\
\cdashline{2-7}
& COS2A & {\bf33.2310} & {\bf2.0312} & {\bf0.0092} & {\bf0.9512} & 14.5959\\
\hline
\multirow{2}{*}{\makecell{Farm}}
& DAC & 27.0976 & 9.0505 & 0.0427 & 0.8896 & \bf{0.0654} \\
\cdashline{2-7}
& COS2A & {\bf35.0685} & {\bf2.1460} & {\bf0.0113} & {\bf0.9527} & 13.9210 \\
\hline
\multirow{2}{*}{\makecell{City}}
& DAC & 28.0807 & 8.4681 & 0.0408 & 0.8230 & \bf{0.0676} \\
\cdashline{2-7}
& COS2A & {\bf35.8443} & {\bf3.3838} & {\bf0.0175} & {\bf0.9209} & 14.6907 \\
\hline
\hline
\multirow{2}{*}{\makecell{Average}}
& DAC & 27.7916 & 7.7147 & 0.0320 & 0.8932 &  \bf{0.0664}\\
\cdashline{2-7}
& COS2A & \bf{35.0084} & \bf{2.5037} & \bf{0.0115} & \bf{0.9488} & 14.6357 \\
\hline
\end{tabular}}
\end{table}

\begin{figure*}[htbp]
\centerline{\includegraphics[width=0.99\textwidth]{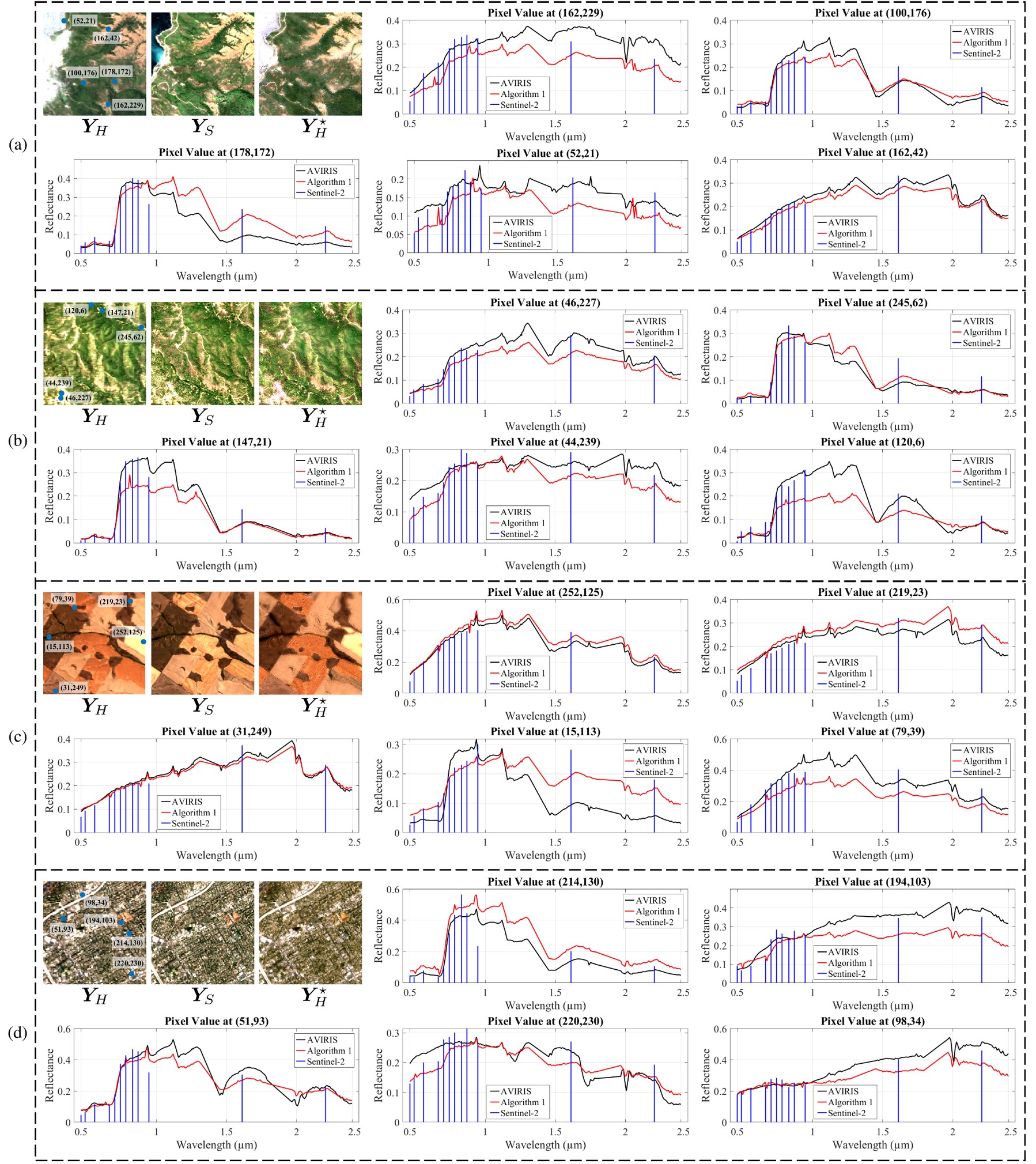}}
\caption{The real data testing results over (a) coastline, (b) mountain, (c) farm, and (d) city.
For each land type, the real AVIRIS image $\bY_H$, its counterpart real Sentinel-2 image $\bY_S$, and the reconstructed AVIRIS image $\bY_H^\star$ (returned by Algorithm \ref{alg:DuQuCODE}) are displayed for visual comparison, where both ($\bY_H$,$\bY_H^\star$) use true-color composition [B23(r)-B12(g)-B5(b)].
The Sentinel-2 image $\bY_S$ also uses the true-color composition [B4(r)-B3(g)-B2(b)].
%=====================
For each land type, we also randomly select five representative pixels with high spectral diversity, and display their Sentinel-2 pixels (as blue pulses), real AVIRIS pixels (as black curves), and reconstructed AVIRIS pixels (as red curves).
}\label{fig:RealDataExp}
\end{figure*}

\subsection{Real Data Evaluation}\label{sec: real result}

This section aims to demonstrate the real-world applicability of the proposed COS2A algorithm (i.e., Algorithm \ref{alg:DuQuCODE}).
Specifically, by directly feeding real Sentinel-2 data $\bY_S$ into the algorithm, we compare the algorithmic output $\bY_H^\star$ with the real AVIRIS counterpart data, thereby showing their high spatial and spectral resemblances.
The preparation of the counterpart AVIRIS data has been reported in Section \ref{sec:experimental setting}.
The other training strategy also follows the same settings as described in Section \ref{sec:experimental setting}.
To effectively evaluate the similarity between $\bY_H^\star$ (generated from real Sentinel-2 data $\bY_S$) and the real AVIRIS data, we perform illumination calibration between the real AVIRIS and real Sentinel-2 data.
Mathematically, the pixel $\ba\in\mathbb{R}^{172}$ in a real AVIRIS image is calibrated as ``$\gamma^\star\ba$'' to fit its counterpart Sentinel-2 pixel $\bs\in\mathbb{R}^{12}$ for some calibration scaler $\gamma^\star\geq 0$, defined as
\begin{equation*}\label{prob:scaled GT}
\gamma^\star
\triangleq 
\arg\min_{\gamma\geq 0}~\frac{1}{2}\|\gamma\widetilde{\ba}-\bs\|_2^2,
\end{equation*} 
where $\widetilde{\ba}\in\mathbb{R}^{12}$ (a subvector of $\ba$) is formed by the 12 AVIRIS bands whose central wavelengths are closest to those of the 12 Sentinel-2 bands.

The calibrated real AVIRIS hyperspectral pixels (i.e., $\gamma^\star\ba$) are displayed as black curves serving as the ground-truth pixels (that can be understood as the desired optimal outcomes of the proposed COS2A algorithm), as can be seen from Figure \ref{fig:RealDataExp}.
Their corresponding COS2A-generated pixels in $\bY_H^\star$ (returned by Algorithm \ref{alg:DuQuCODE}) are then displayed as red curves in Figure \ref{fig:RealDataExp}.
Also, their counterpart real Sentinel-2 pixels (i.e., $\bs$) are displayed as the discrete blue pulses in Figure \ref{fig:RealDataExp}.
We comprehensively investigate four representative land types (i.e., coastline, mountain, farm, and city).
For each land type, the real AVIRIS image $\bY_H$, its counterpart real Sentinel-2 image $\bY_S$, and the reconstructed AVIRIS image $\bY_H^\star$ (returned by Algorithm \ref{alg:DuQuCODE}) are also displayed in Figure \ref{fig:RealDataExp} for visual comparison.
For each land type, we also randomly select five representative pixels with high spectral diversity, and display their multispectral/hyperspectral curves in Figure \ref{fig:RealDataExp}.

For coastline area (cf. Figure \ref{fig:RealDataExp}(a)), one can see that the spatial texture of $\bY_H^\star$ does hold high resemblance to that of $\bY_H$.
It is remarkable that although the coastal area of the $\bY_S$ (the upper-left corner) has some color distortion, our proposed COS2A algorithm is able to automatically calibrate the color, thereby making $\bY_H^\star$ quite similar to $\bY_H$.
One can also see that the real and reconstructed hyperspectral pixels (i.e., the black and red curves) are very similar.
For the pixel at ($178$,$172$), the red curve returned by our COS2A algorithm even better fits the Sentinel-2 multispectral pixel when comparing to the black curve, probably because the acquisition of AVIRIS data is somewhat affected by the illumination condition.
%=======================
For mountain area (cf. Figure \ref{fig:RealDataExp}(b)), the spatial details of $\bY_H^\star$ again hold high resemblance to the real AVIRIS data $\bY_H$, and COS2A-generated hyperspectral pixels are also well reconstructed from $\bY_S$.
%=======================
For farm area (cf. Figure \ref{fig:RealDataExp}(c)), even if $\bY_S$ has obvious illumination deviation from $\bY_H$, the super-resolved $\bY_H^\star$ very well preserves the color and structure of the real AVIRIS data $\bY_H$, preserving clear boundary between different farmlands.
The reconstructed hyperspectral pixels (red curves) also exhibit high fidelity compared to  the real pixels (black curves).
Remarkably, even the Sentinel-2 pixel indexed by ($15$,$113$) is quite deviated in the NIR region, the proposed COS2A algorithm can still return a hyperspectral pixel with very similar spectral shape comparing to the real AVIRIS pixel in the NIR spectral range.
%=======================
For city area (cf. Figure \ref{fig:RealDataExp}(d)), the fine structural information in $\bY_S$ has been well fused into the $\bY_H^\star$, exhibiting consistent and faithful spatial reconstruction performance of the COS2A algorithm. 
The spectral super-resolution is also well achieved with highly aligned real and reconstructed AVIRIS hyperspectral pixels.
%=======================
%
Therefore, we have proven that the proposed COS2A algorithm is able to yield high-fidelity hyperspectral
reconstruction results that hold good resemblance to real AVIRIS data.

%fig caption: done

%fig quality check: still refining

%fig layout illustration: done

%fig discussion (4 land types; 6 similar black/red curves for each type; false color composition is similar to the GT): done

\section{Conclusions and Future Works}\label{sec:conclusion}

We have proposed the COS2A algorithm (i.e., Algorithm \ref{alg:DuQuCODE}) based on a lightweight deep unfolding network (cf. Algorithm \ref{alg:ADMM}) and the spectral-spatial duality (cf. Theorem \ref{thm:SpeSpaDual}), thereby allowing Sentinel-2 multi-resolution data to be computationally upgrade to AVIRIS-level hyperspectral data with uniformly high resolution (i.e., 10 m).
We employ the CODE theory to mitigate the burden of the deep learning, thereby yielding pleasant COS2A results without sophisticated network architecture or laborious data collection procedure.
Theories developed in this paper are not only interpretable, but also can be used to design critical spectral super-resolution algorithms for computationally generating other popular satellite images in the future.
Experiments demonstrate the superiority and effectiveness of our proposed COS2A theory over various land cover types.
In the future, remote sensing researchers could also use our COS2A technique to achieve hyperspectral missions (especially those requiring strong remote substances/objects identifiability) over regions covered by the Sentinel-2 satellite.
Quantum-based real-time software implementation (e.g., by quantum generative models \cite{HyperKING}) and energy-efficient hardware implementation of the COS2A algorithm also deserve future investigations to meet the edge-computing era.

\appendix

\subsection{Proof of Theorem \ref{thm:SpeSpaDual}}\label{sec:proof-thmSSD}

To rigorously build the spectral-spatial duality, we first recall the well-known coupled-NMF spatial super-resolution problem in optical remote sensing \cite{CNMF}, whose formally regularized version can be cast as \cite[Equation (5)]{COCNMF}, which is copied here for the self-contained purpose:
\begin{equation}\label{prob:oriCOCNMF}
\min_{\bA,\bS\geq\bm 0}
\frac{1}{2}\textrm{CNMF}(\bA,\bS)
+\lambda_1 \textrm{vol}(\bA)
+\lambda_2 \|\bS\|_1,
\end{equation}
where $\textrm{CNMF}(\bA,\bS)\triangleq\|\bY_H-\bA\bS\bB\|_F^2+\|\bY_M-\bD\bA\bS\|_F^2$ is the so-called coupled NMF formulation \cite{CNMF} used to spatially super-resolve the hyperspectral image $\bY_H$ by fusing the spatial details of the multispectral image $\bY_M$ into $\bY_H$.
Therefore, it suffices to prove that \eqref{prob:oriCOCNMF} is mathematically equivalent to the target COS2A spectral super-resolution problem \eqref{prob:CA2SE-exp}.

By comparing \eqref{prob:CA2SE-exp} and \eqref{prob:oriCOCNMF}, one can first standardize the parameter setting as $\lambda_1:=0.5\alpha$, $\lambda_2:=0.5\beta$, and $\lambda:=2$ (cf. \eqref{prob:CA2SE-exp}).
Also, the high-resolution multispectral image $\bY_M$ for providing spatial details in the spatial super-resolution problem \eqref{prob:oriCOCNMF} can be set as the image composed of those 10-m Sentinel-2 bands, leading to $\bY_M:=\widetilde{\bY}_S$, and thus $\bD$ in \eqref{prob:oriCOCNMF} is accordingly set as the spectral response $\widetilde{\bD}$ used in \eqref{prob:CA2SE-exp}.
Therefore, in order to prove the equivalence between \eqref{prob:CA2SE-exp} and \eqref{prob:oriCOCNMF}, we only need to prove that the $\bQ$-quadratic-norm term in \eqref{prob:CA2SE-exp} is equivalent to the first term in $\textrm{CNMF}(\bA,\bS)$, i.e., $\|\bY_H-\bA\bS\bB\|_F^2=\|\bA\bS-\bY_\textrm{DE}\|_\bQ^2$.
Furthermore, recalling from the discussion above \eqref{def:Q}, the $r\times r$ region integration (modeled by $\bB$) is first performed on $\bY_\textrm{DE}$ before it is used for $\bQ$-regularization, i.e., $\bY_\textrm{DE}\bB=\bY_H$, implying that the equivalence between \eqref{prob:CA2SE-exp} and \eqref{prob:oriCOCNMF} can be built by proving 
\begin{equation}\label{eq:FisQ}
\|\bY_\textrm{DE}\bB-\bA\bS\bB\|_F^2=\|\bA\bS-\bY_\textrm{DE}\|_\bQ^2,
\end{equation}
which will be proven true below for building the desired spectral-spatial duality.

To prove \eqref{eq:FisQ}, we notice that $\textrm{vec}(\bC\bE\bF)=(\bF^T\otimes\bC)\textrm{vec}(\bE)$ is true for any matrices $(\bC,\bE,\bF)$ with appropriate dimensionality, where $\textrm{vec}(\cdot)$ is the vectorization operator.
Also, by recalling the definition of $\bB\triangleq \bI_{L/r^2}\otimes (\bm 1_{r^2}/r^2)\in \mathbb{R}^{L\times (L/r^2)}$, we can build the following equalities, i.e.,
\begin{align}
&\|\bY_\textrm{DE}\bB-\bA\bS\bB\|_F^2
\notag 
\\
=&\|(\bY_\textrm{DE}-\bA\bS)~\! [\bI_{L/r^2}\otimes (\bm 1_{r^2}/r^2)]\|_F^2
\notag 
\\
=&\|\textrm{vec}\{\bI_M (\bY_\textrm{DE}-\bA\bS)~\! [\bI_{L/r^2}\otimes (\bm 1_{r^2}/r^2)] \}\|_2^2
\notag 
\\
=&\| \{  [\bI_{L/r^2}\otimes (\bm 1_{r^2}/r^2)]^T \otimes \bI_M \} \textrm{vec}(\bY_\textrm{DE}-\bA\bS) \|_2^2. 
\label{eq:AA}
\end{align}
To simplify the derivation, let $\bv\triangleq \textrm{vec}(\bY_\textrm{DE}-\bA\bS)$.
We can then derive that
\begin{align}
&\|\bA\bS-\bY_\textrm{DE}\|_\bQ^2
\notag 
\\
=&
\|\textrm{vec}(\bY_\textrm{DE}-\bA\bS)\|_\bQ^2
\notag 
\\
=& 
\textrm{vec}(\bY_\textrm{DE}-\bA\bS)^T
\bQ 
\textrm{vec}(\bY_\textrm{DE}-\bA\bS)
\notag 
\\
=& 
\textrm{vec}(\bY_\textrm{DE}-\bA\bS)^T
(\bB\bB^T\otimes\bI_M)
\textrm{vec}(\bY_\textrm{DE}-\bA\bS)
\notag 
\\
=& 
\bv^T
(\bB^T\otimes\bI_M)^T
(\bB^T\otimes\bI_M)
\bv
\notag 
\\
=& 
\|(\bB^T\otimes\bI_M)\bv\|_2^2,\label{eq:BB}
\end{align}
where we have used the definition of $\bQ=\bB\bB^T\otimes\bI_M\in\mathbb{R}^{(ML)\times (ML)}$ in \eqref{def:Q}.
By the definitions of $(\bB,\bv)$, we further notice that \eqref{eq:AA} is equivalent to \eqref{eq:BB}.
In other words, $\|\bY_\textrm{DE}\bB-\bA\bS\bB\|_F^2=\|\bA\bS-\bY_\textrm{DE}\|_\bQ^2$, which is exactly \eqref{eq:FisQ}.
Therefore, the spectral-spatial duality is proven, and the proof of Theorem \ref{thm:SpeSpaDual} is completed.

As a side contribution of this proof, we notice that \eqref{prob:CA2SE-exp} is equivalent to \cite[Equation (5)]{COCNMF}, and hence \eqref{prob:CA2SE-exp} can be implemented by calling the coupled NMF algorithm \cite[Algorithm 1]{COCNMF} with the input hyperspectral image set as $\bY_\textrm{DE}\bB$, input multispectral image set as $\widetilde{\bY}_S$, and input parameters empirically set as $(\lambda_1,\lambda_2)=0.5(\alpha,\beta)=(0.001,0.001)$.
\hfill$\blacksquare$

\bibliography{ref}

\begin{IEEEbiography}[{\resizebox{0.9in}{!}{\includegraphics[width=1in,height=1.25in,clip,keepaspectratio]{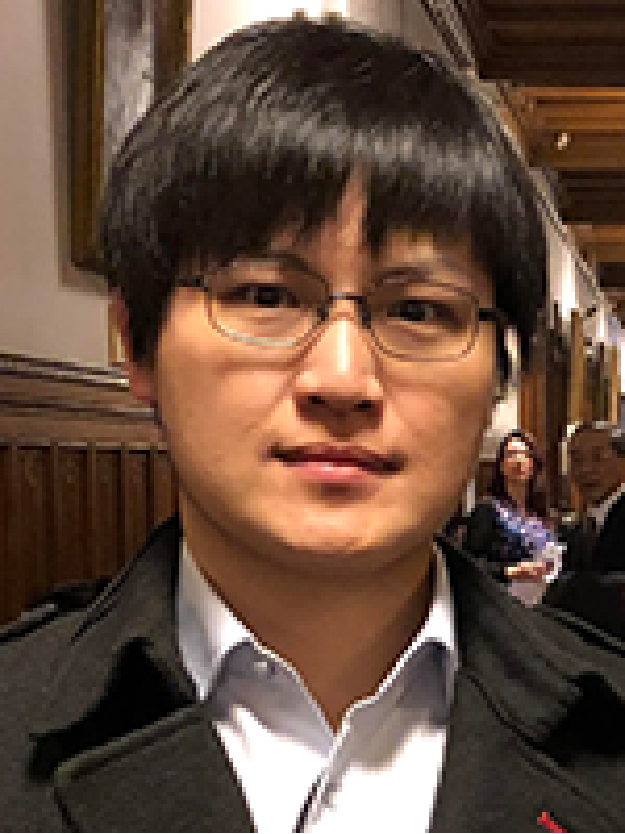}}}]
{\bf Chia-Hsiang Lin}
(S'10-M'18-SM'24)
received the B.S. degree in electrical engineering and the Ph.D. degree in communications engineering from National Tsing Hua University (NTHU), Taiwan, in 2010 and 2016, respectively.
From 2015 to 2016, he was a Visiting Student of Virginia Tech,
Arlington, VA, USA.

He is currently an Associate Professor with the Department of Electrical Engineering, and also with
the Miin Wu School of Computing,
National Cheng Kung University (NCKU), Taiwan.
Before joining NCKU, he held research positions with The Chinese University of Hong Kong, HK (2014 and 2017),
NTHU (2016-2017),
and the University of Lisbon (ULisboa), Lisbon, Portugal (2017-2018).
He was an Assistant Professor with the Center for Space and Remote Sensing Research, National Central University, Taiwan, in 2018, and a Visiting Professor with ULisboa, in 2019.
His research interests include network science,
quantum computing,
convex geometry and optimization, blind signal processing, and imaging science.

Dr. Lin received the Emerging Young Scholar Award (The 2030 Cross-Generation Program) from National Science and Technology Council (NSTC), from 2023 to 2027,
the Future Technology Award from NSTC, in 2022,
the Outstanding Youth Electrical Engineer Award from The Chinese Institute of Electrical Engineering (CIEE), in 2022,
the Best Young Professional Member Award from IEEE Tainan Section, in 2021,
the Prize Paper Award from IEEE Geoscience and Remote Sensing Society (GRS-S), in 2020,
the Top Performance Award from Social Media Prediction Challenge at ACM Multimedia, in 2020,
and The 3rd Place from AIM Real World Super-Resolution Challenge at IEEE International Conference on Computer Vision (ICCV), in 2019.
He received the Ministry of Science and Technology (MOST) Young Scholar Fellowship, together with the EINSTEIN Grant Award, from 2018 to 2023.
In 2016, he was a recipient of the Outstanding Doctoral Dissertation Award from the Chinese Image Processing and Pattern Recognition Society and the Best Doctoral Dissertation Award from the IEEE GRS-S.
\end{IEEEbiography}

\begin{IEEEbiography}[{\resizebox{1in}{!}{\includegraphics[width=1in,height=1.25in,clip,keepaspectratio]{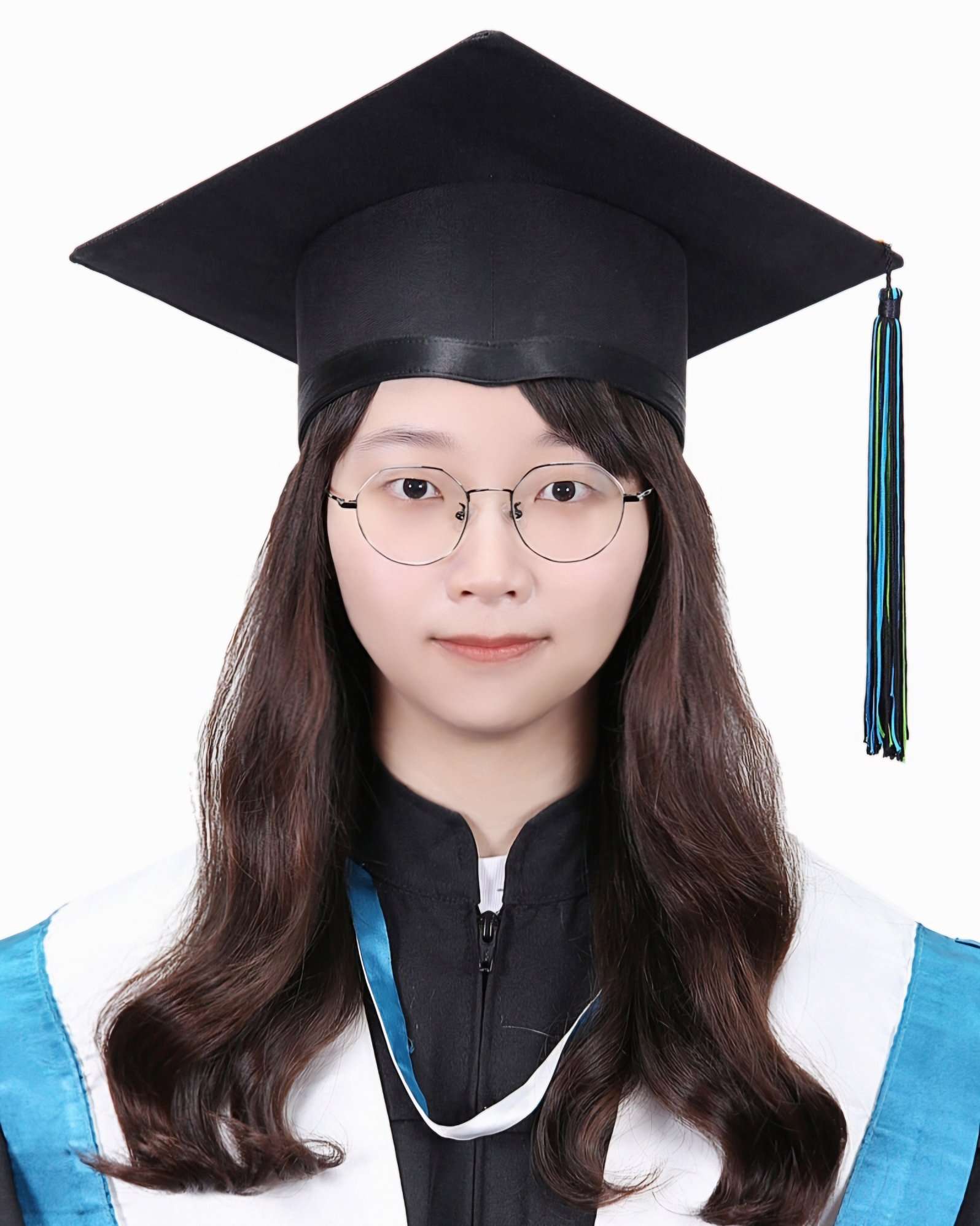}}}]
{\bf Jui-Ting Chen}
received the B.S. degree from the Department of Electrical Engineering, National Sun Yat-Sen University, Kaohsiung, Taiwan, in 2024.

She is currently pursuing the M.S. degree with the Intelligent Hyperspectral Computing Laboratory, Institute of Computer and Communication Engineering, National Cheng Kung University, Tainan, Taiwan.
Her research interests include convex optimization, deep learning, and hyperspectral imaging.
\end{IEEEbiography}

\begin{IEEEbiography}[{\resizebox{1in}{!}{\includegraphics[width=1in,height=1.25in,clip,keepaspectratio]{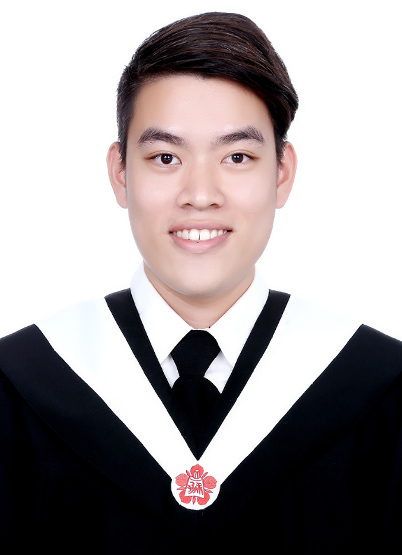}}}]
{\bf Zi-Chao Leng}
received his B.S. degree from the Department of Electronic Engineering, National Cheng Kung University, Taiwan, in 2021.

He is currently a Ph.D. student with Intelligent Hyperspectral Computing Laboratory, Institute of Computer and Communication Engineering, National Cheng Kung University, Taiwan. 
His research interests include deep learning, convex optimization, hyperspectral imaging, and biomedical imaging.
\end{IEEEbiography}

\begin{IEEEbiography}[{\resizebox{1in}{!}{\includegraphics[width=1in,height=1.25in,clip,keepaspectratio]{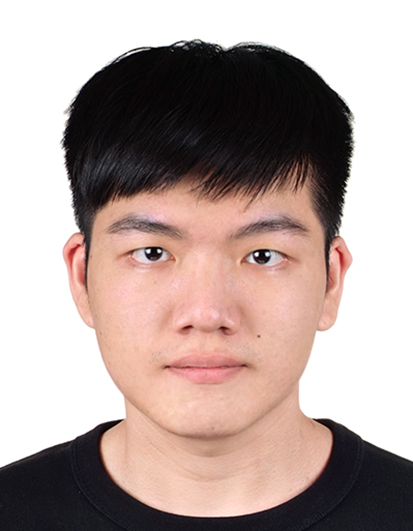}}}]
    	{\bf Jhao-Ting Lin}
		(S'20)
received his B.S. degree from the Department of Communications, Navigation and Control Engineering, National Taiwan Ocean University, Taiwan, in 2020.

He is currently a Ph.D. student affiliated with the Intelligent Hyperspectral Computing Laboratory, Institute of Computer and Communication Engineering, Department of Electrical Engineering, National Cheng Kung University, Taiwan. 
His research interests include convex optimization, deep learning, signal processing, quantum computing, and hyperspectral imaging.
He has received some highly competitive student awards, including the 2022 and 2024 Pan Wen Yuan Award from the Industrial Technology Research Institute (ITRI), Taiwan.
He has been selected as a recipient for the Ph.D. Students Study Abroad Program from the National Science and Technology Council (NSTC), Taiwan, for visiting the Okinawa Institute of Science and Technology in 2025.
\end{IEEEbiography}

\end{document}